\documentclass[twocolumn,superscriptaddress,nofootinbib,amsmath,amssymb,prc]{revtex4-2}

\usepackage{xcolor}
\usepackage{graphicx}
\usepackage{dcolumn}
\usepackage{bm}
\usepackage{ifpdf}
\usepackage{epstopdf}
\usepackage{slashed}
\usepackage{amsfonts}
\usepackage{mathrsfs}
\usepackage{amssymb}
\usepackage{multirow}
\usepackage{CJK}
\usepackage{xcolor}
\usepackage{subfigure,dcolumn}
\usepackage{times}
\usepackage[T2A,T1]{fontenc}
\usepackage[russian,english]{babel}
\usepackage{amsmath}
\usepackage{listings}
\usepackage{booktabs}
\usepackage{color}
\usepackage[misc]{ifsym}
\usepackage{float}
\usepackage{hyperref}

\usepackage{physics}

\makeatletter
\def\NAT@def@citea{\def\@citea{\NAT@separator}}
\makeatother

\begin{document}

\title{Concentrated valence nucleons transfer in heavy-ion collisions: implications for questing the stable superheavy elements}

\author{Zepeng Gao}
\affiliation{Sino-French Institute of Nuclear Engineering and Technology, Sun Yat-sen University, Zhuhai 519082, China}

 \author{Yinu Zhang}
\affiliation{Sino-French Institute of Nuclear Engineering and Technology, Sun Yat-sen University, Zhuhai 519082, China}

\author{Long Zhu}
\email[Contact author, ]{zhulong@mail.sysu.edu.cn}
\affiliation{Sino-French Institute of Nuclear Engineering and Technology, Sun Yat-sen University, Zhuhai 519082, China}
\affiliation{Guangxi Key Laboratory of Nuclear Physics and Nuclear Technology, Guangxi Normal University, Guilin 541004, China}

\date{\today}

\begin{abstract}

The multinucleon transfer process is regarded as a promising pathway for producing the stable superheavy elements. However, the underlying mechanism, especially the possible transfer channels for sailing to the ``island of stability'' are poorly known. 
In this work, the time-dependent Hartree-Fock theory is used to investigate the collision dynamics of $^{136}$Xe, $^{198}$Pt, $^{238}$U + $^{238}$U reactions. 
A novel reaction channel of the concentrated valence nucleons (CVN) transferring is found in the collisions heading on the tips of $^{238}$U. 
These nucleons are transferred with relatively short relaxation time and break the symmetry of nucleon exchange in the early reaction stage. In consequence, the mass equilibrium with relaxation time is deviated from the systematic behavior based on the macroscopic-microscopic potential energy surface. The CVN transfer channel shows promising prospect for producing neutron-rich superheavy nuclei. In this case, we also investigated the angular distributions of products from the CVN transfer channel in the reaction $^{238}$U + $^{238}$U with Tip-Side configuration, and the optimal detection angles are predicted. 

\end{abstract}

\maketitle

\textit{Introduction.} Early calculations taking the shell structure into account predicted that the next shell closures would occur at $Z=114$ and $N=184$ \cite{hofmann2000discovery}. The nuclei around this doubly-magic system were predicted to have half-lives approached the age of the universe \cite{smits2024quest}. The great potential of stable superheavy elements (SHEs) would challenge our understanding of
the nuclear physics as well as chemistry \cite{hofmann2000discovery,oganessian2007heaviest}. In the past decades, the isotopes with or near the $Z=114$ has been produced in the ``cold'' and ``hot'' fusion reactions. However, none of these have come close to comprising 184 neutrons.
To enhance the neutron-richness of the products, the radioactive ion beams could be one possible approach. Unfortunately, the beam intensities are far below the  
required, even at the equipments being currently designed for the future. Based on the fact of profound reconstruction of the initial nuclei in deep-inelastic collisions, alternatively, the multinucleon transfer (MNT) process is regarded as a promising pathway for synthesizing the neutron-rich isotopes, particularly in
reaching the “island of stability” \cite{volkov1978deep, PhysRevLett.101.122701,heinz2022nucleosynthesis}. 


Significant efforts have been devoted to examine critical entrance channel conditions to explore the reaction mechanisms and to promote the production of neutron-rich isotopes. 
For example, the isospin equilibration process, driven by the $N/Z$ ratio difference between the projectile and target nuclei, plays a crucial role in deviating from the $\beta$ stability line \cite{PhysRevC.107.014614,PhysRevC.109.044617}. 
Furthermore, the non-monotonic characteristics of the potential energy surface induced by shell effects profoundly influence evolutionary dynamics \cite{zhu2018shell,liao2024shell}. Utilizing the inverse quasi-fission process driven by shell attraction is regarded as a crucial strategy for producing the transactinide nuclei \cite{zagrebaev2013production,PhysRevC.96.064621}.


Despite considering the aforementioned reaction mechanisms, the quest for ascending the stable SHEs remains a severe challenge. For example, the probability is reduced considerably for transferring a plenty of nucleons. A viable approach such as $^{238}$U-induced actinide-based reactions requires the transfer of approximately 40 nucleons \cite{zagrebaev2013production, Li_2022}. Simultaneously, collision partners that undergo substantial nucleon exchanges frequently attain high excitation energies, thereby reducing the survival probabilities \cite{zagrebaev2013production,ZHU2021136226,ZHU2022137113, LI2020135697,YANG2025139318}.  
Actually, in the most of investigations, only the channel of sequential nucleon transfer is considered.
It has been observed experimentally that the two-nucleon transfer is often enhanced relative to the expectations of independent sequential single-nucleon transfer because of pairing interaction \cite{PhysRevC.66.024606}. 
There is also a distinctive mechanism known as cluster transfer (usually $\alpha$) in nuclear reactions initiated by light nuclei near the Coulomb barrier energy region, which is proposed to describe these reactions based on the assumption that the cluster is separated from the projectile and captured as a whole by the target nucleus \cite{magda1992production,gao2024role}. This is primarily driven by the density distribution inhomogeneity caused by clustering effects in light nuclei. 

The heavy nuclei could also exhibit density inhomogeneity due to nucleons occupying distinct orbitals.
As a quantum many-body system, the nucleons in a nucleus are subject to the Pauli blocking, which restricts them to occupy different energy levels. The occupations of high angular momentum orbitals by valence nucleons in open-shell nuclei break the isotropy of the nucleon density distribution, leading to the deformations of the open-shell nuclei \cite{PhysRevLett.85.720}. 
The correlation mentioned above determines the energy states of the nucleons involved in the reaction, consequently influencing the degree of difficulty of nucleon transfer \cite{mayer1985structure,sekizawa2017enhanced}. This properties in heavy nuclei could introduce a new perspective in MNT reactions, as it may involve the transfer of a large number of nucleons while maintaining relatively low excitation energy \cite{zhao2022distinct}. The full microscopic methods such as time-dependent Hartree-Fock (TDHF) and stochastic mean field (SMF) can effectively incorporate this particular effect, yet without detailed analysis \cite{kedziora2010new,PhysRevC.93.054616,WU2022136886,PhysRevC.100.014612,PhysRevC.108.054605,PhysRevC.102.024619,PhysRevC.107.014609,PhysRevC.111.054613,sekizawa2016time,gao2025time,wakhle2014interplay}. In this Letter, we investigate the nucleon transfer dynamics in the collisions based on different asymmetric configurations with $^{238}$U target using TDHF method.

\begin{figure}[htp!]
\centering
\includegraphics[width=.5\textwidth]{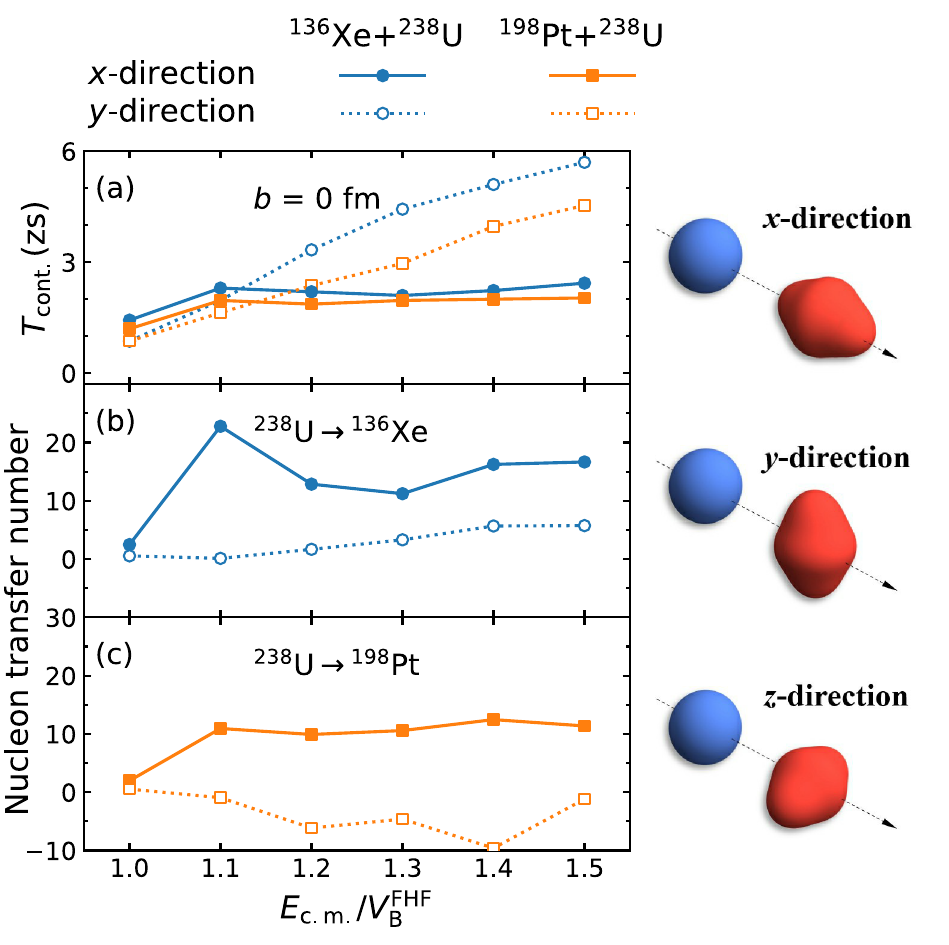}
\caption{\label{fig:1} The reaction contact time (a), the number of the net nucleon transferring from the $^{238}$U to the $^{136}$Xe (b) and $^{198}$Pt (c),  as functions of the reduced incident  center-of-mass energy with impact parameter $b$ = 0 fm. The solid and dotted line represent reactions in the $x$- and $y$-direction, respectively.}
\end{figure}

\textit{Theoretical framework.} 
TDHF is a microscopic dynamical approach that has been extensively employed in the study of low-energy nuclear reactions \cite{RevModPhys.54.913,simenel2012nuclear,SIMENEL201819,sekizawa2019tdhf,STEVENSON2019142, simenel2025nuclearquantummanybodydynamics}. TDHF equation begins with the variation of the action expressed as
$S = \int_{t_1}^{t_2} \langle \Psi(t) | (i\hbar \frac{d}{dt} - \hat{H}) | \Psi(t) \rangle \, dt, $
where \( S \) is the action, \( |\Psi(t)\rangle \) is the many-body wave function, and \(\hat{H}\) is the Hamiltonian. The variation of this action with respect to the wave function \( |\Psi(t)\rangle \) leads to TDHF equation. The wave function \( |\Psi(t)\rangle \) in TDHF framework is typically represented as a single Slater determinant composed of single-particle orbitals:
$|\Psi(t)\rangle = \frac{1}{\sqrt{N!}} \det\{ \psi_i(\boldsymbol{r}\sigma q,t)\}$,
where \(A\) is the number of particles, and $\psi_i(\boldsymbol{r}\sigma q,t)$ is time-dependent single-particle wave functions. 
By performing the variation of the action with respect to single-particle wave functions $\psi_i(\boldsymbol{r}\sigma q,t)$, one derives the equations of motion for the single-particle states. This variation leads to a set of coupled equations, which are essentially the time-dependent Schrödinger equations for each single-particle state in the self-consistent mean field:
$i\hbar \frac{d}{dt} \psi_i(\boldsymbol{r}\sigma q,t) = \hat{h} \psi_i(\boldsymbol{r}\sigma q,t)$,
where \( \hat{h}(t) \) is the single-particle Hamiltonian that includes the mean field generated by all other particles and $\psi$ is single-particle state with spatial coordinate $\boldsymbol{r}$, spin $\sigma$, and isospin $q$. These non-linear equations are solved on a three-dimensional Cartesian grid, enabling a thorough analysis without the imposition of any symmetry constraints. The calculations of static nuclei and dynamic process are described coherently by static HF and TDHF methods using the SKY3D code \cite{schuetrumpf2018tdhf}. The Skyrme density function SLy5 \cite{chabanat1998skyrme} has been utilized in both static and dynamic processes, as its efficacy has been demonstrated in prior quasifission and MNT reactions using TDHF \cite{sekizawa2017enhanced,sekizawa2017microscopic,sekizawa2016time,WU2022136886,PhysRevC.100.014612}. The static HF are employed using the damped gradient iteration method, and the box size was established to be 24 × 24 × 24 fm$^3$ with a mesh spacing of 1.0 fm, while 60 × 24 × 60 fm$^3$ of box size for dynamical simulation was further fixed. 


\textit{Results and discussion.} First, the nucleon transfer dynamics in the mass asymmetric reactions involving $^{136}$Xe and $^{198}$Pt projectiles colliding with prolate ellipsoid deformed $^{238}$U at incident energies ranging from $V_{\rm B}$ to 1.5$V_{\rm B}$ are studied as shown in Fig. \ref{fig:1}. Here, the Frozen Hartree-Fock potential \cite{PhysRevC.95.031601}, incorporating orientation effects in deformed nuclei \cite{PhysRevC.81.064607}, was employed to calculate the Coulomb barrier $V^{\rm FHF}_{\rm B}$ for several reaction systems. Actually, the effect of deformed nuclear orientations in low-energy nuclear reactions has been extensively investigated within the macroscopic models \cite{PhysRevC.103.024608,saiko2019analysis} as well as microscopic transport models \cite{PhysRevC.88.044605,zhang2024study,feng2024microscopic}. However, a continuous and smooth nucleon density distribution is only introduced in above models, which lacks the correlation between the spatial distribution and the occupation of energy levels of nucleons.   
For the central collisions with $b$ = 0 fm, the configuration in the $y$-direction is compact and conducive to the formation of a compound nucleus. Therefore, the lifetime of the dinuclear system (denoted as $T_{\rm cont.}$) increases strongly with the increasing incident energy as shown in Fig. \ref{fig:1}(a). In contrast, $T_{\rm cont.}$ in $x$-direction weakly depends on the incident energy, as the collisions involving the tip orientation usually lead to fast quasifission channels, where enhanced shell effects significantly truncate the interaction duration \cite{wakhle2014interplay}. Interestingly, as shown in Fig. \ref{fig:1}(b) and (c), although the values of $T_{\rm cont.}$ are much higher in $y$-direction, the reactions involving the projectiles of $^{136}$Xe and $^{198}$Pt in the $y$-direction exhibit almost no mass drift towards equilibrium, while significant mass transfer is observed in the $x$-direction. This unexpected behavior that the mass transferring with relaxation time is deviated from the systematic behavior based on the macroscopic-microscopic potential energy surface could be resulted from the inhomogeneous properties in different orientations of the deformed $^{238}$U target. 

\begin{figure}[t!]
\centering
\includegraphics[width=0.4\textwidth]{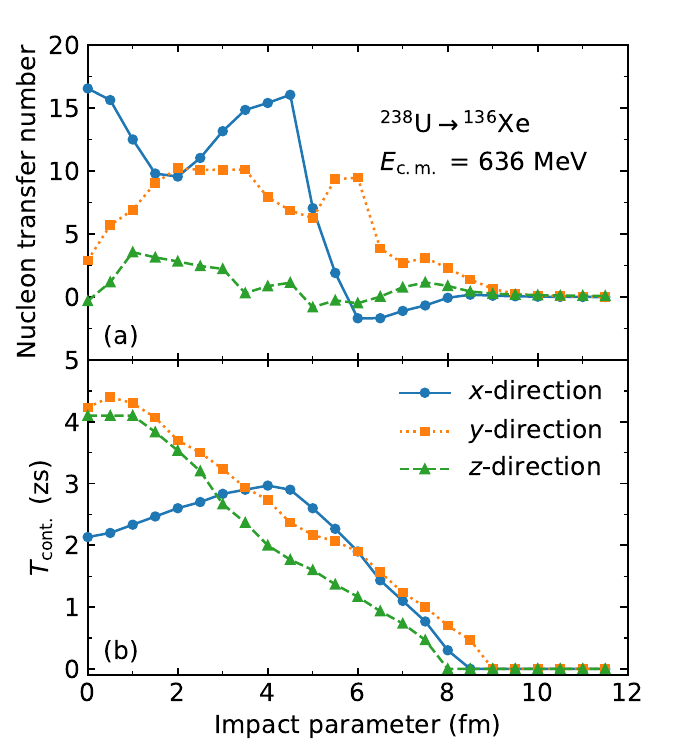}
\caption{\label{fig:2} The net nucleon transfer number from the target to the projectile (a) and the contact time (b) as functions of impact parameter in the reaction of $^{136}$Xe+$^{238}$U at $E_{\rm c.m.}$ = 636 MeV. Solid, dotted, and dashed line denote the results obtained by TDHF with the $x$-, $y$-, and $z$-direction, respectively.
}
\end{figure}

Because of the non-uniform properties in differently oriented configurations of the deformed $^{238}$U target, the dynamics of nucleon transfer could also exhibit significant dependence on the impact parameter.
We further examine the dependence of nucleon transfer and contact time on the impact parameters for the reaction 
$^{136}$Xe + $^{238}$U ($E_{\rm c.m.}$ = 636 MeV) in Fig. \ref{fig:2}. 
For central collisions, a net transfer of 16 nucleons from $^{238}$U to $^{136}$Xe is noticed in collisions along the long axis. However, the collisions along the short axis exhibit almost no mass equilibration process although the relaxation time is about two time longer. 
Notably, the behaviors of nucleon transfer and contact time with impact parameter are quite different between the directions along the long and short axes.
As the impact parameter increases, the configuration and the contact position undergo significant changes. Specifically, for the collisions in the $x$-direction the contact region progressively shifts from tip to side.
One can clearly see that the net transfer of nucleons in the $x$-direction rapidly diminishes when $b>$ 4 fm, whereas a peak is still observed in the $y$-direction for $b=$ 6 fm. 
The crossing of nucleon transfer curves in the $x$- and the $y$-directions is because the reaction regions of tip and side in $^{238}$U interchange as the impact parameters increase. 
In the $z$-direction, nucleons from the short axis of $^{238}$U are consistently involved in the reaction, and this absence of a transition mechanism above results in relatively smooth curves of net nucleon transfer number.

\begin{figure}[htp]
\centering
\includegraphics[width=.48\textwidth]{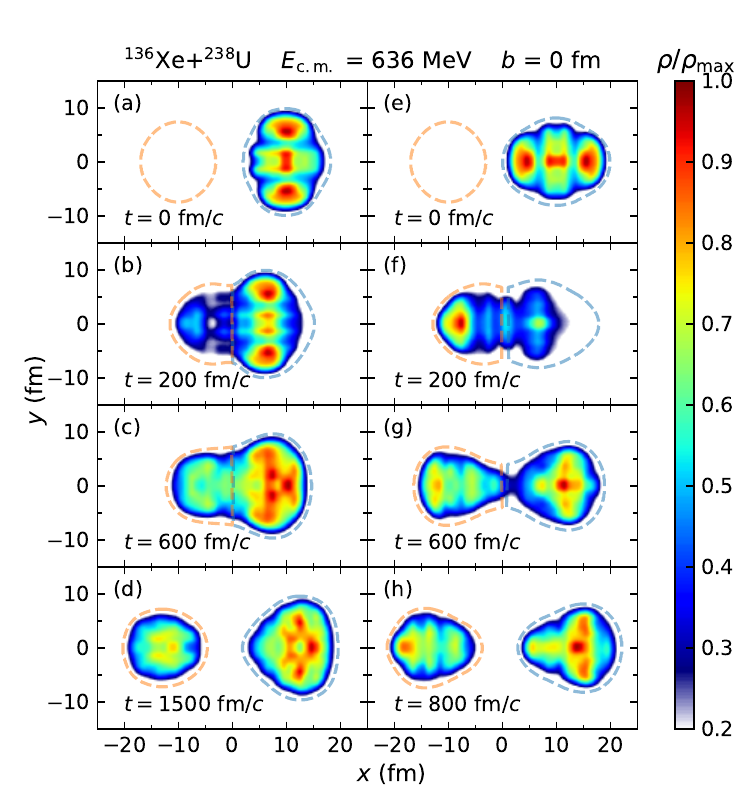}
\caption{\label{fig:4} Time evolution of the normalized two dimensional density distributions of twenty valence nucleons of $^{238}$U for $x$- and $y$-direction reactions of $^{136}$Xe+$^{238}$U at $E_{\rm c.m.}$ = 636 MeV, $b$ = 0 fm. The dashed line of isodensity corresponds to $\rho=0.03$ fm$^{-3}$.}
\end{figure}

To further understand the intriguing orientation effects on the nucleon transfer dynamics, the structural properties of $^{238}$U and the effects on the collision dynamics are investigated. By using the static Hartree-Fock iterations, the ground state of $^{238}$U reveals a clear deformation parameter of $\beta_2 \approx 0.29$, primarily due to the contributions from the valence nucleon density occupancy. The valence nucleons, characterized by high single-particle energies and a tendency to be concentrated at the periphery of the nucleus, are considered more predisposed to transfer during reaction processes. It is hence necessary to discuss the density distributions of the valence nucleons. The normalized density distributions ($\rho/\rho_{\rm max}$) of the valence nucleons obtained from static HF are depicted in Fig. \ref{fig:4}(a) and (e), respectively. Here, 10 neutrons and 10 protons with energies closest to the Fermi surface are chosen as valence nucleons, and the densities are denoted as $\rho_{(20)}(\boldsymbol{r})=\sum_{i\leqslant10}\sum_{\sigma}\sum_{q}\left|\psi_{i}(\boldsymbol{r}\sigma q)\right|^{2}$. The pronounced prolate deformation of $^{238}\text{U}$ is clearly captured. Given the deformation of the nucleus arising from the occupation of high angular momentum orbitals by the valence nucleons, it is evident that a significant portion of the valence neutrons congregate at the tips of $^{238}$U. This creates higher density distributions at the extremities end of the long axis compared to other locations, but this does not mean that the high density at the tips is entirely contributed by the distribution of the valence nucleons.

The valence nucleons are considered to transfer preferentially in nuclear reactions. 
The time evolution of the occupancy distributions of 20 valence nucleons in the projectile and target nuclei is shown in Fig. \ref{fig:4}(a)-(h). The transfer of the valence nucleons in collisions along the $x$-direction becomes apparent at $\sim$200 fm/$c$. Conversely, at this same time point, only a minor fraction of the valence nucleons is transferred in collisions along the short axis, which mainly caused by the diffusion. Additionally, it was observed that valence nucleons exhibit oscillatory motion in the projectile-target collisions along the long axis, which could suppress the net transfer quantity. However, the net transfer quantity remains higher than that observed in $y$-direction, despite the twice the contact time in collisions along the short axis.

\begin{figure}[htp]
\centering
\includegraphics[width=0.45\textwidth]{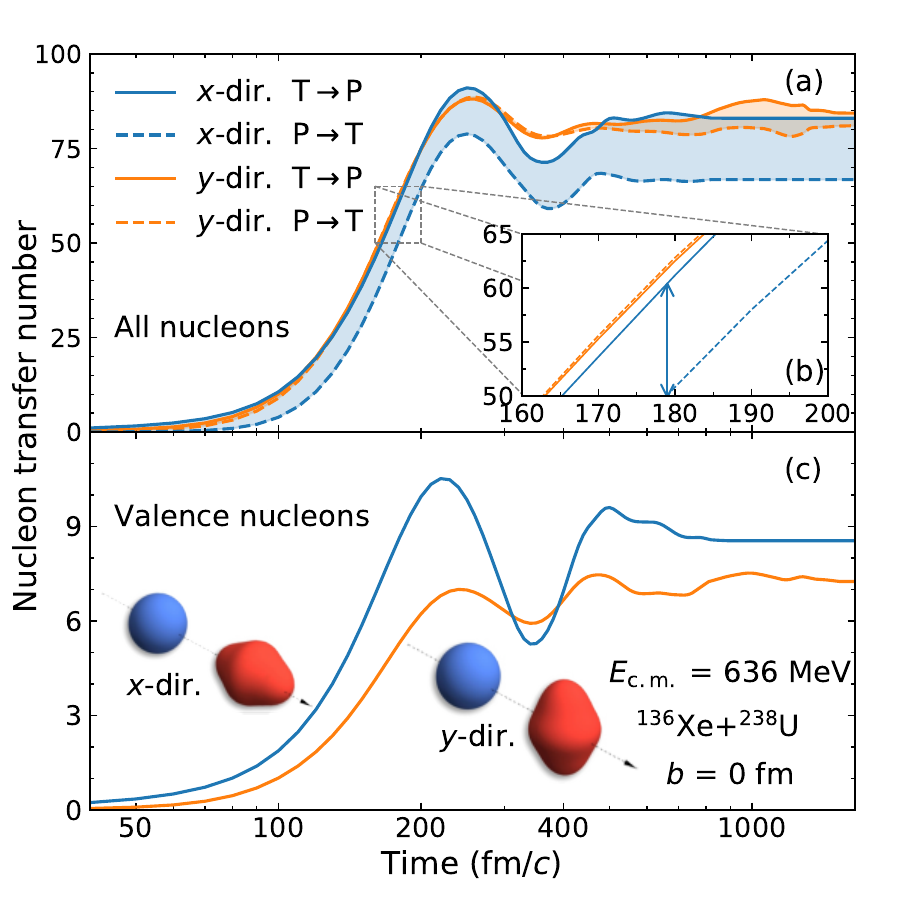}
\caption{\label{fig:5} Time evolution (a logarithmic scale is used) for nucleon transfer number contributed from all nucleons (a) and 20 valence nucleons (c) in the reaction of $^{136}$Xe+$^{238}$U at $E_{\rm c.m.}$ = 636 MeV, $b$ = 0 fm. The blue and orange lines represent the $x$- and $y$-direction, respectively. Solid lines indicate nucleon transfer from $^{238}$U to $^{136}$Xe, while dashed lines denote the inverse transfer. The shaded area illustrates the net transfer number. The time region from 160 fm/$c$ to 200 fm/$c$ is enlarged in (b).}
\end{figure}

An attempt is thus made to extract multiple reaction mechanisms through a more detailed analysis of nucleon transfer. The time evolution for only the valence nucleons transfer number and bidirectional transfer number contributed from all nucleons is shown in Fig. \ref{fig:5}. During the reaction, extensive nucleon exchange occurs, but only a small net transfer number is observed. As shown in Fig. \ref{fig:5}(a), in the $y$-direction, the symmetric nucleon exchange (where the number of nucleon transfer in both directions is approximately the same) occurs within the first 400 fm/$c$, followed by a gradual mass equilibration, as the mass equilibration is a slow process \cite{PhysRevLett.124.212504,yangyu}. However, a significant asymmetric nucleon exchange occurs in the early stage of the reaction in the $x$-direction, with more nucleons being transferred from $^{238}$U to $^{136}$Xe. Typically, a larger window in the $y$-direction could promote more nucleon exchanging between two colliding partners. This is why the nucleon transfer numbers from $^{136}$Xe to $^{238}$U in the $y$-direction exceed those in the $x$-direction, as shown in Fig. \ref{fig:5}(a) and (b). However, interestingly, the transfer enhancement of about 10 nucleons from $^{238}$U to $^{136}$Xe in $x$-direction violates above correlation. If we examine the transfer characteristics of the valence nucleons, we would be able to observe that the nucleon transfer number from $^{238}$U to $^{136}$Xe in the $x$-direction occurs more quickly, with a peak value of 10 nucleons, as shown in Fig. \ref{fig:5}(c). Note that 20 valence nucleons are mainly distributed at two tips of the long axis of $^{238}$U, and the reaction only occurs at one tip, meaning that all the valence nucleons from one tip are transferred to the $^{136}$Xe as one concentrated group. As time evolves, the valence nucleons undergo bidirectional nucleon exchange, resulting in a gradual decrease of the net transfer number. Therefore, we can conclude that in the early stage of the $x$-direction, the valence nucleon transfer occurs, breaking the symmetry of nucleon exchange. The extensive nucleon exchange in the mid-to-late stage suppresses the net transfer of the valence nucleons. However, the effect of the enhanced unidirectional transfer in the early stage persists until the end of the reaction.

\begin{figure}[h!]
\includegraphics[width=0.5\textwidth]{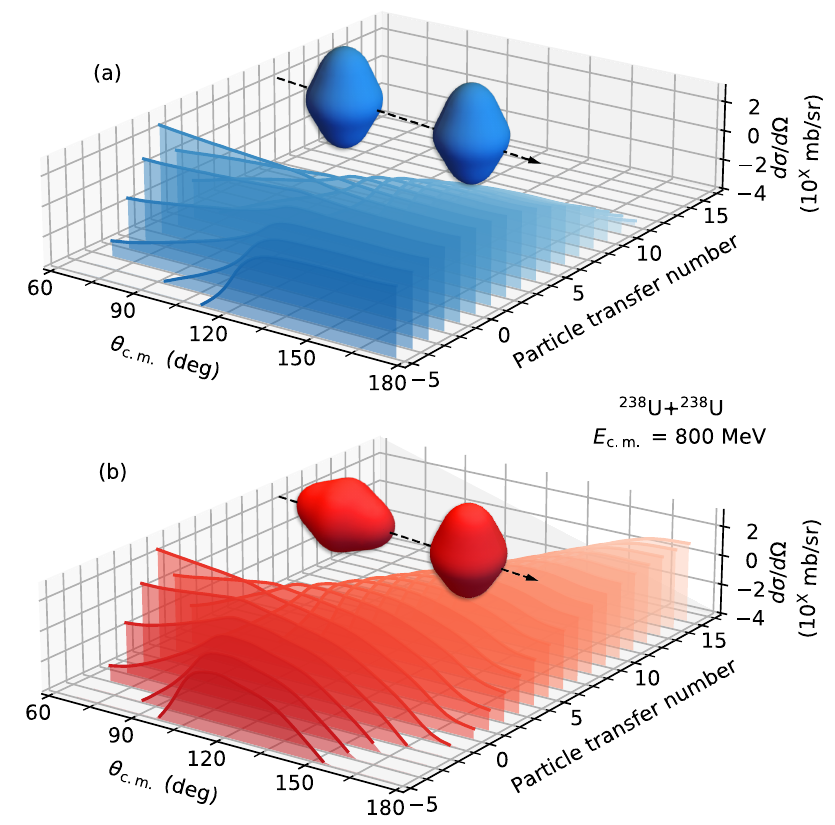}
\caption{\label{fig:6} The angular distributions of primary fragment cross sections for each channel of net transfer particle number in the reaction of $^{238}$U + $^{238}$U at $E_{\rm c.m.}$ = 800 MeV.}
\end{figure}


The observed concentrated valence nucleons (CVN) transfer could facilitate the transfer of a great number of net nucleons with lower excitation energy, which provides potential pathway to explore the ``island of the stability''. Therefore, we finally investigate the collision dynamics of two $^{238}$U atomic nuclei.

Usually, a wide angular distribution limits the measurement of neutron-rich isotopes produced via the MNT process. Also, the non-monotonicity of the nucleon transfer number depended on impact parameters shown in Fig. \ref{fig:2}, caused by the transition of the reaction channels, will be reflected in the angular distributions of the fragments. Therefore, it is essential to find the optimal scattering angle for detecting the products from the CVN transfer channel. The particle number projection (PNP) method \cite{simenel2010particle} was employed to calculate the angular distributions, which is widely used in calculating the cross-sections of MNT reaction products \cite{sekizawa2013time,sekizawa2017microscopic,jiang2020probing,PhysRevC.93.054616,PhysRevC.100.014612,WU2022136886,PhysRevC.109.024614}. The differential cross-sections can be thus calculated with 
$\frac{d\sigma(\theta,N,Z)}{d\Omega}=\frac{b}{\sin\theta}\left|\frac{db}{d\theta(b)}\right|P(b,N,Z)$,
where $\theta$ is scattering angle in the center-of-mass frame, $P(b,N,Z)$ is the probability of product for an isotope with neutron number $N$ and proton number $Z$ in the reaction at an impact parameter $b$ obtained from PNP, and the term of $\left|{db}/{d\theta(b)}\right|$ can be obtained from the deflection function of TDHF calculations. The angular distributions of primary fragment cross sections for each channel of net transfer particle number in the reaction
of $^{238}$U + $^{238}$U at $E_{\rm c.m.}$ = 800 MeV are shown in Fig. \ref{fig:6}, given that actinide-induced MNT reactions are considered a potential pathway to reach the ``island of stability''. The collision orientations of Side-Side and Tip-Side were used to discuss dynamics without and with CVN transfer (the Tip-Tip configuration was also calculated but exhibited ternary quasifission behavior \cite{PhysRevLett.103.042701,PhysRevC.109.024316}), respectively. The angular distributions for each nucleon transfer channel exhibit a monotonic variation in the Side-Side orientation. However, for the Tip-Side configuration, the presence of CVN transfer mechanisms leads to target-like fragments being preferentially emitted at more forward angles  ($\theta_{\rm c.m.} \approx 180$°) as the number of transferred nucleons increases. 

\begin{figure}[h!]
\includegraphics[width=0.45\textwidth]{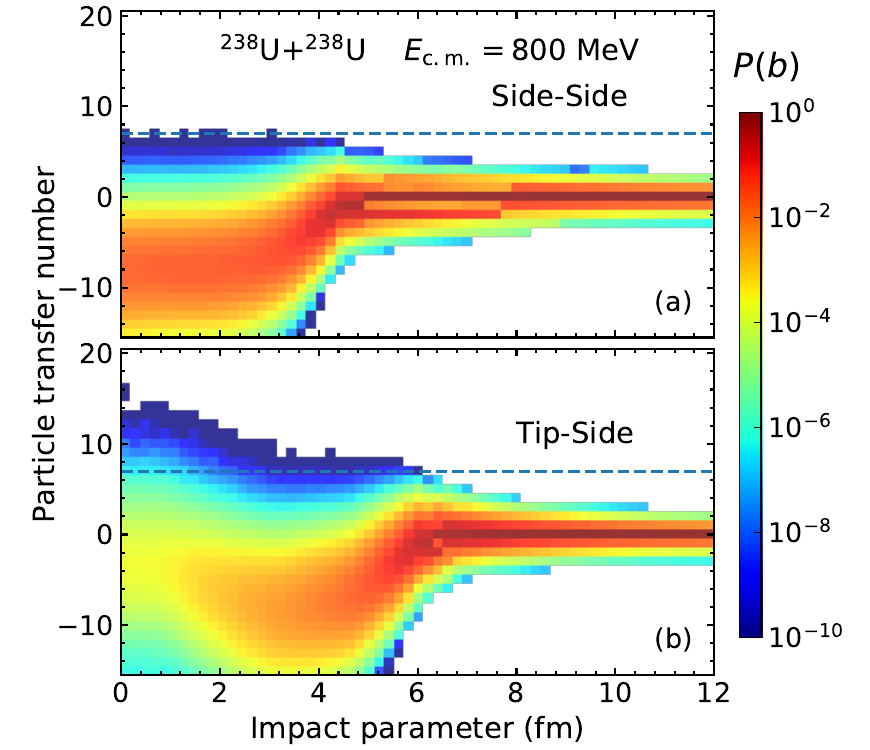}
\caption{\label{fig:7} Particle transfer probabilities of residual target-like fragments in each impact parameter for the reaction of $^{238}$U + $^{238}$U at $E_{\rm c.m.}$ = 800 MeV. The particle transfer number is defined as the difference between the mass number of the target-like fragment after de-excitation and the mass number of $^{238}$U.}
\end{figure}

\begin{figure}[h!]
\includegraphics[width=0.45\textwidth]{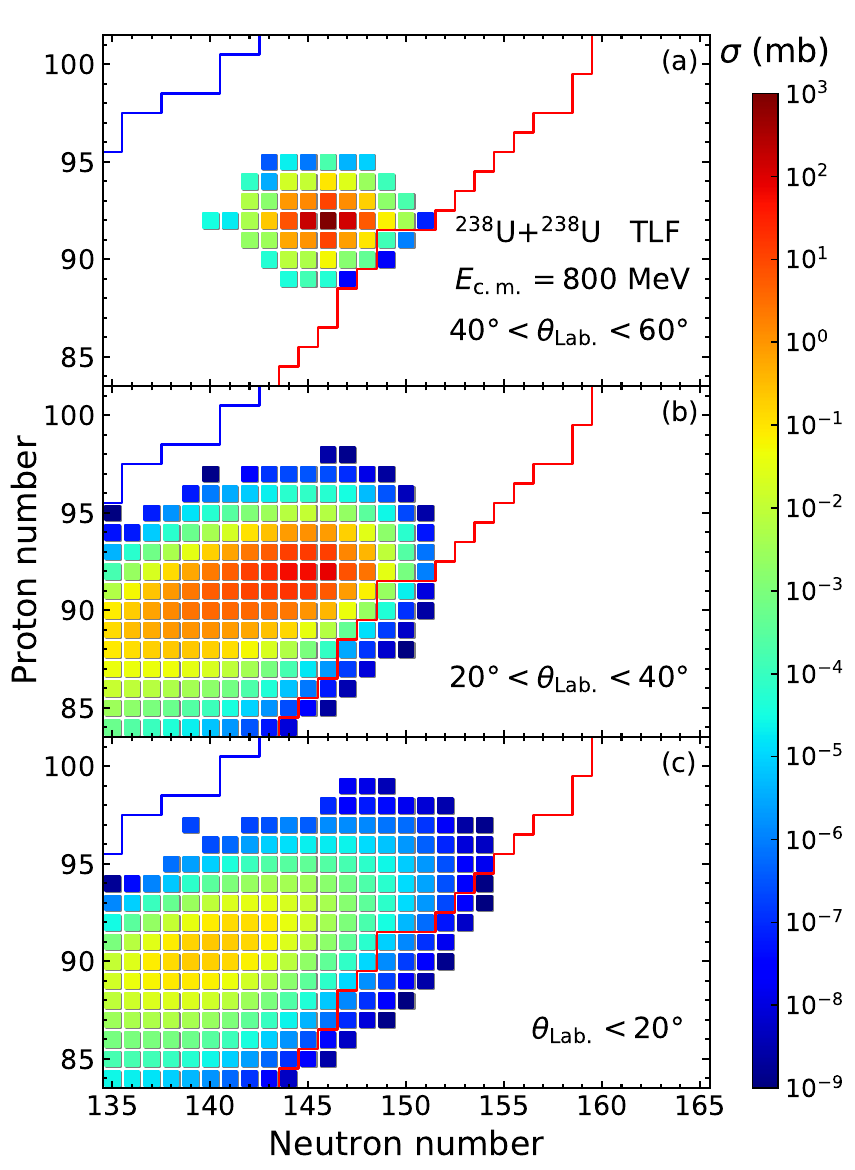}
\caption{\label{fig:8} Isotope distribution of residual target-like fragments in the reaction of $^{238}$U + $^{238}$U at $E_{\rm c.m.}$ = 800 MeV under Tip-Side configurations. Each panel represents different range of emitting angle in the laboratory frame. The blue and red solid lines show the boundaries of known isotopes \cite{boundary}.}
\end{figure}

There was a concern that the forward-angle emitted fragments, resulting from near-central collisions, might exhibit lower yields and be hard to survive under high excitation energies. To address this, we employed the statistical evaporation model GEMINI++ \cite{PhysRevC.82.014610} to conduct 1,000,000 de-excitation simulations for each fragment under each impact parameter, with the results displayed in Fig.~\ref{fig:7}. The $Q$-values are calculated using AME2020 \cite{Wang_2021} masses where available, otherwise the LightGBM-refined FRDM2012 \cite{gao2021machine} predictions. It can be observed that a significant fraction of fragments survive, with some even exhibiting a nucleon transfer number exceeding 10 in near-central collisions under Tip-Side configurations. Finally, the residual cross-sections are plotted on the nuclide chart, with three laboratory angles as shown in Fig.~\ref{fig:8}. The production cross-sections exceeding the nb level are observed in several unknown neutron-rich isotopes within Tip-Side configuration. A larger number of new nuclides tend to be produced at small emission angles, which correspond to near-central collisions where the contribution from valence nucleon transfer becomes more pronounced. This finding theoretically reinforces the feasibility of employing forward-angle detection setups (\textit{e.g.} SHIP \cite{devaraja2020new,devaraja2015observation,devaraja2019population} and TASCA \cite{di2018study}), for probing MNT reactions involving transactinide isotopes \cite{YANG2025139318}.

\textit{Conclusions.} The phenomenon of valence nucleon transfer in heavy-ion reactions involving $^{238}$U as the target nucleus above the Coulomb barrier has been investigated using time-dependent Hartree-Fock approach. In the ground state of $^{238}$U, valence nucleons near the Fermi surface tend to concentrate at the tips of the long axis, which facilitates rapid transfer of these nucleons along this axis. In the early stage of the $x$-direction collisions, the CVN transfer occurs, breaking the symmetry of nucleon exchange. The effect of the enhanced unidirectional transfer in the early stage persists until the end of the reaction. This behavior, characterized by shorter reaction times and large nucleon transfer when $^{238}$U is involved in reactions along its long axis, has been confirmed across various reaction systems, energies, and impact parameters. The angular distributions of products resulting from CVN transfer are analyzed in the collisions of two $^{238}$U atomic nuclei. The small angle emitted target-like fragments exhibit a greater probability of producing unknown neutron-rich heavy nuclides, as the contribution of CVN transfer becomes more pronounced in this region. 

\textit{Acknowledgments.}  The authors thank L. Guo for useful discussions. This work was supported by the National Natural Science Foundation of China under Grants No. 12075327; Fundamental Research Funds for the Central Universities, Sun Yat-sen University under Grant No. 23lgbj003. The authors are grateful to the C3S2 computing center in Huzhou University for calculation support. 


\begin{thebibliography}{67}%
\makeatletter
\providecommand \@ifxundefined [1]{%
 \@ifx{#1\undefined}
}%
\providecommand \@ifnum [1]{%
 \ifnum #1\expandafter \@firstoftwo
 \else \expandafter \@secondoftwo
 \fi
}%
\providecommand \@ifx [1]{%
 \ifx #1\expandafter \@firstoftwo
 \else \expandafter \@secondoftwo
 \fi
}%
\providecommand \natexlab [1]{#1}%
\providecommand \enquote  [1]{``#1''}%
\providecommand \bibnamefont  [1]{#1}%
\providecommand \bibfnamefont [1]{#1}%
\providecommand \citenamefont [1]{#1}%
\providecommand \href@noop [0]{\@secondoftwo}%
\providecommand \href [0]{\begingroup \@sanitize@url \@href}%
\providecommand \@href[1]{\@@startlink{#1}\@@href}%
\providecommand \@@href[1]{\endgroup#1\@@endlink}%
\providecommand \@sanitize@url [0]{\catcode `\\12\catcode `\$12\catcode `\&12\catcode `\#12\catcode `\^12\catcode `\_12\catcode `\%12\relax}%
\providecommand \@@startlink[1]{}%
\providecommand \@@endlink[0]{}%
\providecommand \url  [0]{\begingroup\@sanitize@url \@url }%
\providecommand \@url [1]{\endgroup\@href {#1}{\urlprefix }}%
\providecommand \urlprefix  [0]{URL }%
\providecommand \Eprint [0]{\href }%
\providecommand \doibase [0]{http://dx.doi.org/}%
\providecommand \selectlanguage [0]{\@gobble}%
\providecommand \bibinfo  [0]{\@secondoftwo}%
\providecommand \bibfield  [0]{\@secondoftwo}%
\providecommand \translation [1]{[#1]}%
\providecommand \BibitemOpen [0]{}%
\providecommand \bibitemStop [0]{}%
\providecommand \bibitemNoStop [0]{.\EOS\space}%
\providecommand \EOS [0]{\spacefactor3000\relax}%
\providecommand \BibitemShut  [1]{\csname bibitem#1\endcsname}%
\let\auto@bib@innerbib\@empty
\bibitem [{\citenamefont {Hofmann}\ and\ \citenamefont {M{\"u}nzenberg}(2000)}]{hofmann2000discovery}%
  \BibitemOpen
  \bibfield  {author} {\bibinfo {author} {\bibfnamefont {S.}~\bibnamefont {Hofmann}}\ and\ \bibinfo {author} {\bibfnamefont {G.}~\bibnamefont {M{\"u}nzenberg}},\ }\href {\doibase https://doi.org/10.1103/RevModPhys.72.733} {\bibfield  {journal} {\bibinfo  {journal} {Rev. Mod. Phys.}\ }\textbf {\bibinfo {volume} {72}},\ \bibinfo {pages} {733} (\bibinfo {year} {2000})}\BibitemShut {NoStop}%
\bibitem [{\citenamefont {Smits}\ \emph {et~al.}(2024)\citenamefont {Smits}, \citenamefont {D{\"u}llmann}, \citenamefont {Indelicato}, \citenamefont {Nazarewicz},\ and\ \citenamefont {Schwerdtfeger}}]{smits2024quest}%
  \BibitemOpen
  \bibfield  {author} {\bibinfo {author} {\bibfnamefont {O.~R.}\ \bibnamefont {Smits}}, \bibinfo {author} {\bibfnamefont {C.~E.}\ \bibnamefont {D{\"u}llmann}}, \bibinfo {author} {\bibfnamefont {P.}~\bibnamefont {Indelicato}}, \bibinfo {author} {\bibfnamefont {W.}~\bibnamefont {Nazarewicz}}, \ and\ \bibinfo {author} {\bibfnamefont {P.}~\bibnamefont {Schwerdtfeger}},\ }\href {\doibase https://doi.org/10.1038/s42254-023-00668-y} {\bibfield  {journal} {\bibinfo  {journal} {Nat. Rev. Phys.}\ }\textbf {\bibinfo {volume} {6}},\ \bibinfo {pages} {86} (\bibinfo {year} {2024})}\BibitemShut {NoStop}%
\bibitem [{\citenamefont {Oganessian}(2007)}]{oganessian2007heaviest}%
  \BibitemOpen
  \bibfield  {author} {\bibinfo {author} {\bibfnamefont {Y.}~\bibnamefont {Oganessian}},\ }\href {\doibase 10.1088/0954-3899/34/4/R01} {\bibfield  {journal} {\bibinfo  {journal} {J. Phys. G}\ }\textbf {\bibinfo {volume} {34}},\ \bibinfo {pages} {R165} (\bibinfo {year} {2007})}\BibitemShut {NoStop}%
\bibitem [{\citenamefont {Volkov}(1978)}]{volkov1978deep}%
  \BibitemOpen
  \bibfield  {author} {\bibinfo {author} {\bibfnamefont {V.~V.}\ \bibnamefont {Volkov}},\ }\href {\doibase https://doi.org/10.1016/0370-1573(78)90200-4} {\bibfield  {journal} {\bibinfo  {journal} {Physics Reports}\ }\textbf {\bibinfo {volume} {44}},\ \bibinfo {pages} {93} (\bibinfo {year} {1978})}\BibitemShut {NoStop}%
\bibitem [{\citenamefont {Zagrebaev}\ and\ \citenamefont {Greiner}(2008)}]{PhysRevLett.101.122701}%
  \BibitemOpen
  \bibfield  {author} {\bibinfo {author} {\bibfnamefont {V.}~\bibnamefont {Zagrebaev}}\ and\ \bibinfo {author} {\bibfnamefont {W.}~\bibnamefont {Greiner}},\ }\href {\doibase 10.1103/PhysRevLett.101.122701} {\bibfield  {journal} {\bibinfo  {journal} {Phys. Rev. Lett.}\ }\textbf {\bibinfo {volume} {101}},\ \bibinfo {pages} {122701} (\bibinfo {year} {2008})}\BibitemShut {NoStop}%
\bibitem [{\citenamefont {Heinz}\ and\ \citenamefont {Devaraja}(2022)}]{heinz2022nucleosynthesis}%
  \BibitemOpen
  \bibfield  {author} {\bibinfo {author} {\bibfnamefont {S.}~\bibnamefont {Heinz}}\ and\ \bibinfo {author} {\bibfnamefont {H.}~\bibnamefont {Devaraja}},\ }\href {\doibase https://doi.org/10.1140/epja/s10050-022-00771-1} {\bibfield  {journal} {\bibinfo  {journal} {Eur. Phys. J. A}\ }\textbf {\bibinfo {volume} {58}},\ \bibinfo {pages} {114} (\bibinfo {year} {2022})}\BibitemShut {NoStop}%
\bibitem [{\citenamefont {Liao}\ \emph {et~al.}(2023)\citenamefont {Liao}, \citenamefont {Zhu}, \citenamefont {Su},\ and\ \citenamefont {Li}}]{PhysRevC.107.014614}%
  \BibitemOpen
  \bibfield  {author} {\bibinfo {author} {\bibfnamefont {Z.}~\bibnamefont {Liao}}, \bibinfo {author} {\bibfnamefont {L.}~\bibnamefont {Zhu}}, \bibinfo {author} {\bibfnamefont {J.}~\bibnamefont {Su}}, \ and\ \bibinfo {author} {\bibfnamefont {C.}~\bibnamefont {Li}},\ }\href {\doibase 10.1103/PhysRevC.107.014614} {\bibfield  {journal} {\bibinfo  {journal} {Phys. Rev. C}\ }\textbf {\bibinfo {volume} {107}},\ \bibinfo {pages} {014614} (\bibinfo {year} {2023})}\BibitemShut {NoStop}%
\bibitem [{\citenamefont {Zhang}\ \emph {et~al.}(2024{\natexlab{a}})\citenamefont {Zhang}, \citenamefont {Li}, \citenamefont {Li}, \citenamefont {Zhang}, \citenamefont {Zou},\ and\ \citenamefont {Zhang}}]{PhysRevC.109.044617}%
  \BibitemOpen
  \bibfield  {author} {\bibinfo {author} {\bibfnamefont {Y.-H.}\ \bibnamefont {Zhang}}, \bibinfo {author} {\bibfnamefont {J.-J.}\ \bibnamefont {Li}}, \bibinfo {author} {\bibfnamefont {C.}~\bibnamefont {Li}}, \bibinfo {author} {\bibfnamefont {M.-H.}\ \bibnamefont {Zhang}}, \bibinfo {author} {\bibfnamefont {Y.}~\bibnamefont {Zou}}, \ and\ \bibinfo {author} {\bibfnamefont {F.-S.}\ \bibnamefont {Zhang}},\ }\href {\doibase 10.1103/PhysRevC.109.044617} {\bibfield  {journal} {\bibinfo  {journal} {Phys. Rev. C}\ }\textbf {\bibinfo {volume} {109}},\ \bibinfo {pages} {044617} (\bibinfo {year} {2024}{\natexlab{a}})}\BibitemShut {NoStop}%
\bibitem [{\citenamefont {Zhu}\ \emph {et~al.}(2018)\citenamefont {Zhu}, \citenamefont {Wen}, \citenamefont {Lin}, \citenamefont {Bao}, \citenamefont {Su}, \citenamefont {Li},\ and\ \citenamefont {Guo}}]{zhu2018shell}%
  \BibitemOpen
  \bibfield  {author} {\bibinfo {author} {\bibfnamefont {L.}~\bibnamefont {Zhu}}, \bibinfo {author} {\bibfnamefont {P.-W.}\ \bibnamefont {Wen}}, \bibinfo {author} {\bibfnamefont {C.-J.}\ \bibnamefont {Lin}}, \bibinfo {author} {\bibfnamefont {X.-J.}\ \bibnamefont {Bao}}, \bibinfo {author} {\bibfnamefont {J.}~\bibnamefont {Su}}, \bibinfo {author} {\bibfnamefont {C.}~\bibnamefont {Li}}, \ and\ \bibinfo {author} {\bibfnamefont {C.-C.}\ \bibnamefont {Guo}},\ }\href {\doibase 10.1103/PhysRevC.97.044614} {\bibfield  {journal} {\bibinfo  {journal} {Phys. Rev. C}\ }\textbf {\bibinfo {volume} {97}},\ \bibinfo {pages} {044614} (\bibinfo {year} {2018})}\BibitemShut {NoStop}%
\bibitem [{\citenamefont {Liao}\ \emph {et~al.}(2024)\citenamefont {Liao}, \citenamefont {Gao}, \citenamefont {Yang}, \citenamefont {Zhu},\ and\ \citenamefont {Su}}]{liao2024shell}%
  \BibitemOpen
  \bibfield  {author} {\bibinfo {author} {\bibfnamefont {Z.}~\bibnamefont {Liao}}, \bibinfo {author} {\bibfnamefont {Z.}~\bibnamefont {Gao}}, \bibinfo {author} {\bibfnamefont {Y.}~\bibnamefont {Yang}}, \bibinfo {author} {\bibfnamefont {L.}~\bibnamefont {Zhu}}, \ and\ \bibinfo {author} {\bibfnamefont {J.}~\bibnamefont {Su}},\ }\href {\doibase 10.1103/PhysRevC.109.054612} {\bibfield  {journal} {\bibinfo  {journal} {Phys. Rev. C}\ }\textbf {\bibinfo {volume} {109}},\ \bibinfo {pages} {054612} (\bibinfo {year} {2024})}\BibitemShut {NoStop}%
\bibitem [{\citenamefont {Zagrebaev}\ and\ \citenamefont {Greiner}(2013)}]{zagrebaev2013production}%
  \BibitemOpen
  \bibfield  {author} {\bibinfo {author} {\bibfnamefont {V.~I.}\ \bibnamefont {Zagrebaev}}\ and\ \bibinfo {author} {\bibfnamefont {W.}~\bibnamefont {Greiner}},\ }\href {\doibase 10.1103/PhysRevC.87.034608} {\bibfield  {journal} {\bibinfo  {journal} {Phys. Rev. C}\ }\textbf {\bibinfo {volume} {87}},\ \bibinfo {pages} {034608} (\bibinfo {year} {2013})}\BibitemShut {NoStop}%
\bibitem [{\citenamefont {Kozulin}\ \emph {et~al.}(2017)\citenamefont {Kozulin}, \citenamefont {Zagrebaev}, \citenamefont {Knyazheva}, \citenamefont {Itkis}, \citenamefont {Novikov}, \citenamefont {Itkis}, \citenamefont {Dmitriev}, \citenamefont {Harca}, \citenamefont {Bondarchenko}, \citenamefont {Karpov}, \citenamefont {Saiko},\ and\ \citenamefont {Vardaci}}]{PhysRevC.96.064621}%
  \BibitemOpen
  \bibfield  {author} {\bibinfo {author} {\bibfnamefont {E.~M.}\ \bibnamefont {Kozulin}}, \bibinfo {author} {\bibfnamefont {V.~I.}\ \bibnamefont {Zagrebaev}}, \bibinfo {author} {\bibfnamefont {G.~N.}\ \bibnamefont {Knyazheva}}, \bibinfo {author} {\bibfnamefont {I.~M.}\ \bibnamefont {Itkis}}, \bibinfo {author} {\bibfnamefont {K.~V.}\ \bibnamefont {Novikov}}, \bibinfo {author} {\bibfnamefont {M.~G.}\ \bibnamefont {Itkis}}, \bibinfo {author} {\bibfnamefont {S.~N.}\ \bibnamefont {Dmitriev}}, \bibinfo {author} {\bibfnamefont {I.~M.}\ \bibnamefont {Harca}}, \bibinfo {author} {\bibfnamefont {A.~E.}\ \bibnamefont {Bondarchenko}}, \bibinfo {author} {\bibfnamefont {A.~V.}\ \bibnamefont {Karpov}}, \bibinfo {author} {\bibfnamefont {V.~V.}\ \bibnamefont {Saiko}}, \ and\ \bibinfo {author} {\bibfnamefont {E.}~\bibnamefont {Vardaci}},\ }\href {\doibase 10.1103/PhysRevC.96.064621} {\bibfield  {journal} {\bibinfo  {journal} {Phys. Rev. C}\ }\textbf {\bibinfo {volume} {96}},\ \bibinfo {pages} {064621} (\bibinfo {year}
  {2017})}\BibitemShut {NoStop}%
\bibitem [{\citenamefont {Li}\ \emph {et~al.}(2022)\citenamefont {Li}, \citenamefont {Zhang}, \citenamefont {Zhang}, \citenamefont {Zhang}, \citenamefont {Liu},\ and\ \citenamefont {Zhang}}]{Li_2022}%
  \BibitemOpen
  \bibfield  {author} {\bibinfo {author} {\bibfnamefont {J.}~\bibnamefont {Li}}, \bibinfo {author} {\bibfnamefont {G.}~\bibnamefont {Zhang}}, \bibinfo {author} {\bibfnamefont {X.}~\bibnamefont {Zhang}}, \bibinfo {author} {\bibfnamefont {Y.}~\bibnamefont {Zhang}}, \bibinfo {author} {\bibfnamefont {Z.}~\bibnamefont {Liu}}, \ and\ \bibinfo {author} {\bibfnamefont {F.-S.}\ \bibnamefont {Zhang}},\ }\href {\doibase 10.1088/1361-6471/ac44ad} {\bibfield  {journal} {\bibinfo  {journal} {J. Phys. G}\ }\textbf {\bibinfo {volume} {49}},\ \bibinfo {pages} {025106} (\bibinfo {year} {2022})}\BibitemShut {NoStop}%
\bibitem [{\citenamefont {Zhu}(2021)}]{ZHU2021136226}%
  \BibitemOpen
  \bibfield  {author} {\bibinfo {author} {\bibfnamefont {L.}~\bibnamefont {Zhu}},\ }\href {\doibase https://doi.org/10.1016/j.physletb.2021.136226} {\bibfield  {journal} {\bibinfo  {journal} {Phys. Lett. B}\ }\textbf {\bibinfo {volume} {816}},\ \bibinfo {pages} {136226} (\bibinfo {year} {2021})}\BibitemShut {NoStop}%
\bibitem [{\citenamefont {Zhu}\ \emph {et~al.}(2022)\citenamefont {Zhu}, \citenamefont {Su}, \citenamefont {Li},\ and\ \citenamefont {Zhang}}]{ZHU2022137113}%
  \BibitemOpen
  \bibfield  {author} {\bibinfo {author} {\bibfnamefont {L.}~\bibnamefont {Zhu}}, \bibinfo {author} {\bibfnamefont {J.}~\bibnamefont {Su}}, \bibinfo {author} {\bibfnamefont {C.}~\bibnamefont {Li}}, \ and\ \bibinfo {author} {\bibfnamefont {F.-S.}\ \bibnamefont {Zhang}},\ }\href {\doibase https://doi.org/10.1016/j.physletb.2022.137113} {\bibfield  {journal} {\bibinfo  {journal} {Phys. Lett. B}\ }\textbf {\bibinfo {volume} {829}},\ \bibinfo {pages} {137113} (\bibinfo {year} {2022})}\BibitemShut {NoStop}%
\bibitem [{\citenamefont {Li}\ \emph {et~al.}(2020)\citenamefont {Li}, \citenamefont {Tian},\ and\ \citenamefont {Zhang}}]{LI2020135697}%
  \BibitemOpen
  \bibfield  {author} {\bibinfo {author} {\bibfnamefont {C.}~\bibnamefont {Li}}, \bibinfo {author} {\bibfnamefont {J.}~\bibnamefont {Tian}}, \ and\ \bibinfo {author} {\bibfnamefont {F.-S.}\ \bibnamefont {Zhang}},\ }\href {\doibase https://doi.org/10.1016/j.physletb.2020.135697} {\bibfield  {journal} {\bibinfo  {journal} {Phys. Lett. B}\ }\textbf {\bibinfo {volume} {809}},\ \bibinfo {pages} {135697} (\bibinfo {year} {2020})}\BibitemShut {NoStop}%
\bibitem [{\citenamefont {Yang}\ \emph {et~al.}(2025{\natexlab{a}})\citenamefont {Yang}, \citenamefont {Zhu}, \citenamefont {Liao}, \citenamefont {Gao}, \citenamefont {Fang}, \citenamefont {Su}, \citenamefont {Liu}, \citenamefont {Hou}, \citenamefont {Huang}, \citenamefont {Li} \emph {et~al.}}]{YANG2025139318}%
  \BibitemOpen
  \bibfield  {author} {\bibinfo {author} {\bibfnamefont {Y.}~\bibnamefont {Yang}}, \bibinfo {author} {\bibfnamefont {L.}~\bibnamefont {Zhu}}, \bibinfo {author} {\bibfnamefont {Z.}~\bibnamefont {Liao}}, \bibinfo {author} {\bibfnamefont {Z.}~\bibnamefont {Gao}}, \bibinfo {author} {\bibfnamefont {Y.}~\bibnamefont {Fang}}, \bibinfo {author} {\bibfnamefont {J.}~\bibnamefont {Su}}, \bibinfo {author} {\bibfnamefont {Z.}~\bibnamefont {Liu}}, \bibinfo {author} {\bibfnamefont {D.}~\bibnamefont {Hou}}, \bibinfo {author} {\bibfnamefont {H.}~\bibnamefont {Huang}}, \bibinfo {author} {\bibfnamefont {C.}~\bibnamefont {Li}},  \emph {et~al.},\ }\href {\doibase https://doi.org/10.1016/j.physletb.2025.139318} {\bibfield  {journal} {\bibinfo  {journal} {Phys. Lett. B}\ }\textbf {\bibinfo {volume} {862}},\ \bibinfo {pages} {139318} (\bibinfo {year} {2025}{\natexlab{a}})}\BibitemShut {NoStop}%
\bibitem [{\citenamefont {Corradi}\ \emph {et~al.}(2002)\citenamefont {Corradi}, \citenamefont {Vinodkumar}, \citenamefont {Stefanini}, \citenamefont {Fioretto}, \citenamefont {Prete}, \citenamefont {Beghini}, \citenamefont {Montagnoli}, \citenamefont {Scarlassara}, \citenamefont {Pollarolo}, \citenamefont {Cerutti},\ and\ \citenamefont {Winther}}]{PhysRevC.66.024606}%
  \BibitemOpen
  \bibfield  {author} {\bibinfo {author} {\bibfnamefont {L.}~\bibnamefont {Corradi}}, \bibinfo {author} {\bibfnamefont {A.~M.}\ \bibnamefont {Vinodkumar}}, \bibinfo {author} {\bibfnamefont {A.~M.}\ \bibnamefont {Stefanini}}, \bibinfo {author} {\bibfnamefont {E.}~\bibnamefont {Fioretto}}, \bibinfo {author} {\bibfnamefont {G.}~\bibnamefont {Prete}}, \bibinfo {author} {\bibfnamefont {S.}~\bibnamefont {Beghini}}, \bibinfo {author} {\bibfnamefont {G.}~\bibnamefont {Montagnoli}}, \bibinfo {author} {\bibfnamefont {F.}~\bibnamefont {Scarlassara}}, \bibinfo {author} {\bibfnamefont {G.}~\bibnamefont {Pollarolo}}, \bibinfo {author} {\bibfnamefont {F.}~\bibnamefont {Cerutti}}, \ and\ \bibinfo {author} {\bibfnamefont {A.}~\bibnamefont {Winther}},\ }\href {\doibase 10.1103/PhysRevC.66.024606} {\bibfield  {journal} {\bibinfo  {journal} {Phys. Rev. C}\ }\textbf {\bibinfo {volume} {66}},\ \bibinfo {pages} {024606} (\bibinfo {year} {2002})}\BibitemShut {NoStop}%
\bibitem [{\citenamefont {Magda}\ and\ \citenamefont {Leyba}(1992)}]{magda1992production}%
  \BibitemOpen
  \bibfield  {author} {\bibinfo {author} {\bibfnamefont {M.}~\bibnamefont {Magda}}\ and\ \bibinfo {author} {\bibfnamefont {J.}~\bibnamefont {Leyba}},\ }\href {\doibase https://doi.org/10.1142/S0218301392000114} {\bibfield  {journal} {\bibinfo  {journal} {Int. J. Mod. Phys. E}\ }\textbf {\bibinfo {volume} {1}},\ \bibinfo {pages} {221} (\bibinfo {year} {1992})}\BibitemShut {NoStop}%
\bibitem [{\citenamefont {Gao}\ \emph {et~al.}(2024)\citenamefont {Gao}, \citenamefont {Zhang}, \citenamefont {Zhu}, \citenamefont {Liao}, \citenamefont {Yang}, \citenamefont {Guo},\ and\ \citenamefont {Su}}]{gao2024role}%
  \BibitemOpen
  \bibfield  {author} {\bibinfo {author} {\bibfnamefont {Z.}~\bibnamefont {Gao}}, \bibinfo {author} {\bibfnamefont {Y.}~\bibnamefont {Zhang}}, \bibinfo {author} {\bibfnamefont {L.}~\bibnamefont {Zhu}}, \bibinfo {author} {\bibfnamefont {Z.}~\bibnamefont {Liao}}, \bibinfo {author} {\bibfnamefont {Y.}~\bibnamefont {Yang}}, \bibinfo {author} {\bibfnamefont {C.}~\bibnamefont {Guo}}, \ and\ \bibinfo {author} {\bibfnamefont {J.}~\bibnamefont {Su}},\ }\href {\doibase 10.1103/PhysRevC.109.L041605} {\bibfield  {journal} {\bibinfo  {journal} {Phys. Rev. C}\ }\textbf {\bibinfo {volume} {109}},\ \bibinfo {pages} {L041605} (\bibinfo {year} {2024})}\BibitemShut {NoStop}%
\bibitem [{\citenamefont {Zhao}\ \emph {et~al.}(2000)\citenamefont {Zhao}, \citenamefont {Casten},\ and\ \citenamefont {Arima}}]{PhysRevLett.85.720}%
  \BibitemOpen
  \bibfield  {author} {\bibinfo {author} {\bibfnamefont {Y.~M.}\ \bibnamefont {Zhao}}, \bibinfo {author} {\bibfnamefont {R.~F.}\ \bibnamefont {Casten}}, \ and\ \bibinfo {author} {\bibfnamefont {A.}~\bibnamefont {Arima}},\ }\href {\doibase 10.1103/PhysRevLett.85.720} {\bibfield  {journal} {\bibinfo  {journal} {Phys. Rev. Lett.}\ }\textbf {\bibinfo {volume} {85}},\ \bibinfo {pages} {720} (\bibinfo {year} {2000})}\BibitemShut {NoStop}%
\bibitem [{\citenamefont {Mayer}\ \emph {et~al.}(1985)\citenamefont {Mayer}, \citenamefont {Beier}, \citenamefont {Friese}, \citenamefont {Henning}, \citenamefont {Kienle}, \citenamefont {K{\"o}rner}, \citenamefont {Mayer}, \citenamefont {M{\"u}ller}, \citenamefont {Rosner},\ and\ \citenamefont {Wagner}}]{mayer1985structure}%
  \BibitemOpen
  \bibfield  {author} {\bibinfo {author} {\bibfnamefont {W.}~\bibnamefont {Mayer}}, \bibinfo {author} {\bibfnamefont {G.}~\bibnamefont {Beier}}, \bibinfo {author} {\bibfnamefont {J.}~\bibnamefont {Friese}}, \bibinfo {author} {\bibfnamefont {W.}~\bibnamefont {Henning}}, \bibinfo {author} {\bibfnamefont {P.}~\bibnamefont {Kienle}}, \bibinfo {author} {\bibfnamefont {H.}~\bibnamefont {K{\"o}rner}}, \bibinfo {author} {\bibfnamefont {W.}~\bibnamefont {Mayer}}, \bibinfo {author} {\bibfnamefont {L.}~\bibnamefont {M{\"u}ller}}, \bibinfo {author} {\bibfnamefont {G.}~\bibnamefont {Rosner}}, \ and\ \bibinfo {author} {\bibfnamefont {W.}~\bibnamefont {Wagner}},\ }\href {\doibase https://doi.org/10.1016/0370-2693(85)91161-X} {\bibfield  {journal} {\bibinfo  {journal} {Phys. Lett. B}\ }\textbf {\bibinfo {volume} {152}},\ \bibinfo {pages} {162} (\bibinfo {year} {1985})}\BibitemShut {NoStop}%
\bibitem [{\citenamefont {Sekizawa}(2017{\natexlab{a}})}]{sekizawa2017enhanced}%
  \BibitemOpen
  \bibfield  {author} {\bibinfo {author} {\bibfnamefont {K.}~\bibnamefont {Sekizawa}},\ }\href {\doibase 10.1103/PhysRevC.96.041601} {\bibfield  {journal} {\bibinfo  {journal} {Phys. Rev. C}\ }\textbf {\bibinfo {volume} {96}},\ \bibinfo {pages} {041601(R)} (\bibinfo {year} {2017}{\natexlab{a}})}\BibitemShut {NoStop}%
\bibitem [{\citenamefont {Zhao}\ \emph {et~al.}(2022)\citenamefont {Zhao}, \citenamefont {Liu}, \citenamefont {Zhang}, \citenamefont {Wang},\ and\ \citenamefont {Duan}}]{zhao2022distinct}%
  \BibitemOpen
  \bibfield  {author} {\bibinfo {author} {\bibfnamefont {K.}~\bibnamefont {Zhao}}, \bibinfo {author} {\bibfnamefont {Z.}~\bibnamefont {Liu}}, \bibinfo {author} {\bibfnamefont {F.~S.}\ \bibnamefont {Zhang}}, \bibinfo {author} {\bibfnamefont {N.}~\bibnamefont {Wang}}, \ and\ \bibinfo {author} {\bibfnamefont {J.~Z.}\ \bibnamefont {Duan}},\ }\href {\doibase 10.1103/PhysRevC.106.L011602} {\bibfield  {journal} {\bibinfo  {journal} {Phys. Rev. C}\ }\textbf {\bibinfo {volume} {106}},\ \bibinfo {pages} {L011602} (\bibinfo {year} {2022})}\BibitemShut {NoStop}%
\bibitem [{\citenamefont {Kedziora}\ and\ \citenamefont {Simenel}(2010)}]{kedziora2010new}%
  \BibitemOpen
  \bibfield  {author} {\bibinfo {author} {\bibfnamefont {D.~J.}\ \bibnamefont {Kedziora}}\ and\ \bibinfo {author} {\bibfnamefont {C.}~\bibnamefont {Simenel}},\ }\href {\doibase 10.1103/PhysRevC.81.044613} {\bibfield  {journal} {\bibinfo  {journal} {Phys. Rev. C}\ }\textbf {\bibinfo {volume} {81}},\ \bibinfo {pages} {044613} (\bibinfo {year} {2010})}\BibitemShut {NoStop}%
\bibitem [{\citenamefont {Sekizawa}\ and\ \citenamefont {Yabana}(2016{\natexlab{a}})}]{PhysRevC.93.054616}%
  \BibitemOpen
  \bibfield  {author} {\bibinfo {author} {\bibfnamefont {K.}~\bibnamefont {Sekizawa}}\ and\ \bibinfo {author} {\bibfnamefont {K.}~\bibnamefont {Yabana}},\ }\href {\doibase 10.1103/PhysRevC.93.054616} {\bibfield  {journal} {\bibinfo  {journal} {Phys. Rev. C}\ }\textbf {\bibinfo {volume} {93}},\ \bibinfo {pages} {054616} (\bibinfo {year} {2016}{\natexlab{a}})}\BibitemShut {NoStop}%
\bibitem [{\citenamefont {Wu}\ \emph {et~al.}(2022)\citenamefont {Wu}, \citenamefont {Guo}, \citenamefont {Liu},\ and\ \citenamefont {Peng}}]{WU2022136886}%
  \BibitemOpen
  \bibfield  {author} {\bibinfo {author} {\bibfnamefont {Z.}~\bibnamefont {Wu}}, \bibinfo {author} {\bibfnamefont {L.}~\bibnamefont {Guo}}, \bibinfo {author} {\bibfnamefont {Z.}~\bibnamefont {Liu}}, \ and\ \bibinfo {author} {\bibfnamefont {G.}~\bibnamefont {Peng}},\ }\href {\doibase https://doi.org/10.1016/j.physletb.2022.136886} {\bibfield  {journal} {\bibinfo  {journal} {Phys. Lett. B}\ }\textbf {\bibinfo {volume} {825}},\ \bibinfo {pages} {136886} (\bibinfo {year} {2022})}\BibitemShut {NoStop}%
\bibitem [{\citenamefont {Wu}\ and\ \citenamefont {Guo}(2019)}]{PhysRevC.100.014612}%
  \BibitemOpen
  \bibfield  {author} {\bibinfo {author} {\bibfnamefont {Z.}~\bibnamefont {Wu}}\ and\ \bibinfo {author} {\bibfnamefont {L.}~\bibnamefont {Guo}},\ }\href {\doibase 10.1103/PhysRevC.100.014612} {\bibfield  {journal} {\bibinfo  {journal} {Phys. Rev. C}\ }\textbf {\bibinfo {volume} {100}},\ \bibinfo {pages} {014612} (\bibinfo {year} {2019})}\BibitemShut {NoStop}%
\bibitem [{\citenamefont {Ayik}\ \emph {et~al.}(2023{\natexlab{a}})\citenamefont {Ayik}, \citenamefont {Arik}, \citenamefont {Erbayri}, \citenamefont {Yilmaz},\ and\ \citenamefont {Umar}}]{PhysRevC.108.054605}%
  \BibitemOpen
  \bibfield  {author} {\bibinfo {author} {\bibfnamefont {S.}~\bibnamefont {Ayik}}, \bibinfo {author} {\bibfnamefont {M.}~\bibnamefont {Arik}}, \bibinfo {author} {\bibfnamefont {E.}~\bibnamefont {Erbayri}}, \bibinfo {author} {\bibfnamefont {O.}~\bibnamefont {Yilmaz}}, \ and\ \bibinfo {author} {\bibfnamefont {A.~S.}\ \bibnamefont {Umar}},\ }\href {\doibase 10.1103/PhysRevC.108.054605} {\bibfield  {journal} {\bibinfo  {journal} {Phys. Rev. C}\ }\textbf {\bibinfo {volume} {108}},\ \bibinfo {pages} {054605} (\bibinfo {year} {2023}{\natexlab{a}})}\BibitemShut {NoStop}%
\bibitem [{\citenamefont {Ayik}\ \emph {et~al.}(2020)\citenamefont {Ayik}, \citenamefont {Yilmaz}, \citenamefont {Yilmaz},\ and\ \citenamefont {Umar}}]{PhysRevC.102.024619}%
  \BibitemOpen
  \bibfield  {author} {\bibinfo {author} {\bibfnamefont {S.}~\bibnamefont {Ayik}}, \bibinfo {author} {\bibfnamefont {B.}~\bibnamefont {Yilmaz}}, \bibinfo {author} {\bibfnamefont {O.}~\bibnamefont {Yilmaz}}, \ and\ \bibinfo {author} {\bibfnamefont {A.~S.}\ \bibnamefont {Umar}},\ }\href {\doibase 10.1103/PhysRevC.102.024619} {\bibfield  {journal} {\bibinfo  {journal} {Phys. Rev. C}\ }\textbf {\bibinfo {volume} {102}},\ \bibinfo {pages} {024619} (\bibinfo {year} {2020})}\BibitemShut {NoStop}%
\bibitem [{\citenamefont {Ayik}\ \emph {et~al.}(2023{\natexlab{b}})\citenamefont {Ayik}, \citenamefont {Arik}, \citenamefont {Yilmaz}, \citenamefont {Yilmaz},\ and\ \citenamefont {Umar}}]{PhysRevC.107.014609}%
  \BibitemOpen
  \bibfield  {author} {\bibinfo {author} {\bibfnamefont {S.}~\bibnamefont {Ayik}}, \bibinfo {author} {\bibfnamefont {M.}~\bibnamefont {Arik}}, \bibinfo {author} {\bibfnamefont {O.}~\bibnamefont {Yilmaz}}, \bibinfo {author} {\bibfnamefont {B.}~\bibnamefont {Yilmaz}}, \ and\ \bibinfo {author} {\bibfnamefont {A.~S.}\ \bibnamefont {Umar}},\ }\href {\doibase 10.1103/PhysRevC.107.014609} {\bibfield  {journal} {\bibinfo  {journal} {Phys. Rev. C}\ }\textbf {\bibinfo {volume} {107}},\ \bibinfo {pages} {014609} (\bibinfo {year} {2023}{\natexlab{b}})}\BibitemShut {NoStop}%
\bibitem [{\citenamefont {Ocal}\ \emph {et~al.}(2025)\citenamefont {Ocal}, \citenamefont {Yilmaz}, \citenamefont {Arik}, \citenamefont {Ayik},\ and\ \citenamefont {Umar}}]{PhysRevC.111.054613}%
  \BibitemOpen
  \bibfield  {author} {\bibinfo {author} {\bibfnamefont {S.~E.}\ \bibnamefont {Ocal}}, \bibinfo {author} {\bibfnamefont {O.}~\bibnamefont {Yilmaz}}, \bibinfo {author} {\bibfnamefont {M.}~\bibnamefont {Arik}}, \bibinfo {author} {\bibfnamefont {S.}~\bibnamefont {Ayik}}, \ and\ \bibinfo {author} {\bibfnamefont {A.~S.}\ \bibnamefont {Umar}},\ }\href {\doibase 10.1103/PhysRevC.111.054613} {\bibfield  {journal} {\bibinfo  {journal} {Phys. Rev. C}\ }\textbf {\bibinfo {volume} {111}},\ \bibinfo {pages} {054613} (\bibinfo {year} {2025})}\BibitemShut {NoStop}%
\bibitem [{\citenamefont {Sekizawa}\ and\ \citenamefont {Yabana}(2016{\natexlab{b}})}]{sekizawa2016time}%
  \BibitemOpen
  \bibfield  {author} {\bibinfo {author} {\bibfnamefont {K.}~\bibnamefont {Sekizawa}}\ and\ \bibinfo {author} {\bibfnamefont {K.}~\bibnamefont {Yabana}},\ }\href {\doibase 10.1103/PhysRevC.93.054616} {\bibfield  {journal} {\bibinfo  {journal} {Phys. Rev. C}\ }\textbf {\bibinfo {volume} {93}},\ \bibinfo {pages} {054616} (\bibinfo {year} {2016}{\natexlab{b}})}\BibitemShut {NoStop}%
\bibitem [{\citenamefont {Gao}\ \emph {et~al.}(2025)\citenamefont {Gao}, \citenamefont {Sekizawa},\ and\ \citenamefont {Zhu}}]{gao2025time}%
  \BibitemOpen
  \bibfield  {author} {\bibinfo {author} {\bibfnamefont {Z.}~\bibnamefont {Gao}}, \bibinfo {author} {\bibfnamefont {K.}~\bibnamefont {Sekizawa}}, \ and\ \bibinfo {author} {\bibfnamefont {L.}~\bibnamefont {Zhu}},\ }\href {\doibase 10.1103/zz3y-22fh} {\bibfield  {journal} {\bibinfo  {journal} {Phys. Rev. C}\ }\textbf {\bibinfo {volume} {112}},\ \bibinfo {pages} {014602} (\bibinfo {year} {2025})}\BibitemShut {NoStop}%
\bibitem [{\citenamefont {Wakhle}\ \emph {et~al.}(2014)\citenamefont {Wakhle}, \citenamefont {Simenel}, \citenamefont {Hinde}, \citenamefont {Dasgupta}, \citenamefont {Evers}, \citenamefont {Luong}, \citenamefont {du~Rietz},\ and\ \citenamefont {Williams}}]{wakhle2014interplay}%
  \BibitemOpen
  \bibfield  {author} {\bibinfo {author} {\bibfnamefont {A.}~\bibnamefont {Wakhle}}, \bibinfo {author} {\bibfnamefont {C.}~\bibnamefont {Simenel}}, \bibinfo {author} {\bibfnamefont {D.~J.}\ \bibnamefont {Hinde}}, \bibinfo {author} {\bibfnamefont {M.}~\bibnamefont {Dasgupta}}, \bibinfo {author} {\bibfnamefont {M.}~\bibnamefont {Evers}}, \bibinfo {author} {\bibfnamefont {D.~H.}\ \bibnamefont {Luong}}, \bibinfo {author} {\bibfnamefont {R.}~\bibnamefont {du~Rietz}}, \ and\ \bibinfo {author} {\bibfnamefont {E.}~\bibnamefont {Williams}},\ }\href {\doibase 10.1103/PhysRevLett.113.182502} {\bibfield  {journal} {\bibinfo  {journal} {Phys. Rev. Lett.}\ }\textbf {\bibinfo {volume} {113}},\ \bibinfo {pages} {182502} (\bibinfo {year} {2014})}\BibitemShut {NoStop}%
\bibitem [{\citenamefont {Negele}(1982)}]{RevModPhys.54.913}%
  \BibitemOpen
  \bibfield  {author} {\bibinfo {author} {\bibfnamefont {J.~W.}\ \bibnamefont {Negele}},\ }\href {\doibase 10.1103/RevModPhys.54.913} {\bibfield  {journal} {\bibinfo  {journal} {Rev. Mod. Phys.}\ }\textbf {\bibinfo {volume} {54}},\ \bibinfo {pages} {913} (\bibinfo {year} {1982})}\BibitemShut {NoStop}%
\bibitem [{\citenamefont {Simenel}(2012)}]{simenel2012nuclear}%
  \BibitemOpen
  \bibfield  {author} {\bibinfo {author} {\bibfnamefont {C.}~\bibnamefont {Simenel}},\ }\href {\doibase https://doi.org/10.1140/epja/i2012-12152-0} {\bibfield  {journal} {\bibinfo  {journal} {Eur. Phys. J. A}\ }\textbf {\bibinfo {volume} {48}},\ \bibinfo {pages} {152} (\bibinfo {year} {2012})}\BibitemShut {NoStop}%
\bibitem [{\citenamefont {Simenel}\ and\ \citenamefont {Umar}(2018)}]{SIMENEL201819}%
  \BibitemOpen
  \bibfield  {author} {\bibinfo {author} {\bibfnamefont {C.}~\bibnamefont {Simenel}}\ and\ \bibinfo {author} {\bibfnamefont {A.}~\bibnamefont {Umar}},\ }\href {\doibase https://doi.org/10.1016/j.ppnp.2018.07.002} {\bibfield  {journal} {\bibinfo  {journal} {Prog. Part. Nucl. Phys.}\ }\textbf {\bibinfo {volume} {103}},\ \bibinfo {pages} {19} (\bibinfo {year} {2018})}\BibitemShut {NoStop}%
\bibitem [{\citenamefont {Sekizawa}(2019)}]{sekizawa2019tdhf}%
  \BibitemOpen
  \bibfield  {author} {\bibinfo {author} {\bibfnamefont {K.}~\bibnamefont {Sekizawa}},\ }\href {\doibase https://doi.org/10.3389/fphy.2019.00020} {\bibfield  {journal} {\bibinfo  {journal} {Front. Phys.}\ }\textbf {\bibinfo {volume} {7}},\ \bibinfo {pages} {20} (\bibinfo {year} {2019})}\BibitemShut {NoStop}%
\bibitem [{\citenamefont {Stevenson}\ and\ \citenamefont {Barton}(2019)}]{STEVENSON2019142}%
  \BibitemOpen
  \bibfield  {author} {\bibinfo {author} {\bibfnamefont {P.}~\bibnamefont {Stevenson}}\ and\ \bibinfo {author} {\bibfnamefont {M.}~\bibnamefont {Barton}},\ }\href {\doibase https://doi.org/10.1016/j.ppnp.2018.09.002} {\bibfield  {journal} {\bibinfo  {journal} {Prog. Part. Nucl. Phys.}\ }\textbf {\bibinfo {volume} {104}},\ \bibinfo {pages} {142} (\bibinfo {year} {2019})}\BibitemShut {NoStop}%
\bibitem [{\citenamefont {Simenel}(2025)}]{simenel2025nuclearquantummanybodydynamics}%
  \BibitemOpen
  \bibfield  {author} {\bibinfo {author} {\bibfnamefont {C.}~\bibnamefont {Simenel}},\ }\href {https://arxiv.org/abs/2506.04261} {\bibfield  {journal} {\bibinfo  {journal} {arXiv preprint arXiv:2506.04261}\ } (\bibinfo {year} {2025})}\BibitemShut {NoStop}%
\bibitem [{\citenamefont {Schuetrumpf}\ \emph {et~al.}(2018)\citenamefont {Schuetrumpf}, \citenamefont {Reinhard}, \citenamefont {Stevenson}, \citenamefont {Umar},\ and\ \citenamefont {Maruhn}}]{schuetrumpf2018tdhf}%
  \BibitemOpen
  \bibfield  {author} {\bibinfo {author} {\bibfnamefont {B.}~\bibnamefont {Schuetrumpf}}, \bibinfo {author} {\bibfnamefont {P.-G.}\ \bibnamefont {Reinhard}}, \bibinfo {author} {\bibfnamefont {P.}~\bibnamefont {Stevenson}}, \bibinfo {author} {\bibfnamefont {A.~S.}\ \bibnamefont {Umar}}, \ and\ \bibinfo {author} {\bibfnamefont {J.~A.}\ \bibnamefont {Maruhn}},\ }\href {\doibase https://doi.org/10.1016/j.cpc.2018.03.012} {\bibfield  {journal} {\bibinfo  {journal} {Comput. Phys. Commun.}\ }\textbf {\bibinfo {volume} {229}},\ \bibinfo {pages} {211} (\bibinfo {year} {2018})}\BibitemShut {NoStop}%
\bibitem [{\citenamefont {Chabanat}\ \emph {et~al.}(1998)\citenamefont {Chabanat}, \citenamefont {Bonche}, \citenamefont {Haensel}, \citenamefont {Meyer},\ and\ \citenamefont {Schaeffer}}]{chabanat1998skyrme}%
  \BibitemOpen
  \bibfield  {author} {\bibinfo {author} {\bibfnamefont {E.}~\bibnamefont {Chabanat}}, \bibinfo {author} {\bibfnamefont {P.}~\bibnamefont {Bonche}}, \bibinfo {author} {\bibfnamefont {P.}~\bibnamefont {Haensel}}, \bibinfo {author} {\bibfnamefont {J.}~\bibnamefont {Meyer}}, \ and\ \bibinfo {author} {\bibfnamefont {R.}~\bibnamefont {Schaeffer}},\ }\href {\doibase https://doi.org/10.1016/S0375-9474(98)00180-8} {\bibfield  {journal} {\bibinfo  {journal} {Nucl. Phys. A}\ }\textbf {\bibinfo {volume} {635}},\ \bibinfo {pages} {231} (\bibinfo {year} {1998})}\BibitemShut {NoStop}%
\bibitem [{\citenamefont {Sekizawa}(2017{\natexlab{b}})}]{sekizawa2017microscopic}%
  \BibitemOpen
  \bibfield  {author} {\bibinfo {author} {\bibfnamefont {K.}~\bibnamefont {Sekizawa}},\ }\href {\doibase 10.1103/PhysRevC.96.014615} {\bibfield  {journal} {\bibinfo  {journal} {Phys. Rev. C}\ }\textbf {\bibinfo {volume} {96}},\ \bibinfo {pages} {014615} (\bibinfo {year} {2017}{\natexlab{b}})}\BibitemShut {NoStop}%
\bibitem [{\citenamefont {Simenel}\ \emph {et~al.}(2017)\citenamefont {Simenel}, \citenamefont {Umar}, \citenamefont {Godbey}, \citenamefont {Dasgupta},\ and\ \citenamefont {Hinde}}]{PhysRevC.95.031601}%
  \BibitemOpen
  \bibfield  {author} {\bibinfo {author} {\bibfnamefont {C.}~\bibnamefont {Simenel}}, \bibinfo {author} {\bibfnamefont {A.~S.}\ \bibnamefont {Umar}}, \bibinfo {author} {\bibfnamefont {K.}~\bibnamefont {Godbey}}, \bibinfo {author} {\bibfnamefont {M.}~\bibnamefont {Dasgupta}}, \ and\ \bibinfo {author} {\bibfnamefont {D.~J.}\ \bibnamefont {Hinde}},\ }\href {\doibase 10.1103/PhysRevC.95.031601} {\bibfield  {journal} {\bibinfo  {journal} {Phys. Rev. C}\ }\textbf {\bibinfo {volume} {95}},\ \bibinfo {pages} {031601(R)} (\bibinfo {year} {2017})}\BibitemShut {NoStop}%
\bibitem [{\citenamefont {Umar}\ \emph {et~al.}(2010)\citenamefont {Umar}, \citenamefont {Oberacker}, \citenamefont {Maruhn},\ and\ \citenamefont {Reinhard}}]{PhysRevC.81.064607}%
  \BibitemOpen
  \bibfield  {author} {\bibinfo {author} {\bibfnamefont {A.~S.}\ \bibnamefont {Umar}}, \bibinfo {author} {\bibfnamefont {V.~E.}\ \bibnamefont {Oberacker}}, \bibinfo {author} {\bibfnamefont {J.~A.}\ \bibnamefont {Maruhn}}, \ and\ \bibinfo {author} {\bibfnamefont {P.-G.}\ \bibnamefont {Reinhard}},\ }\href {\doibase 10.1103/PhysRevC.81.064607} {\bibfield  {journal} {\bibinfo  {journal} {Phys. Rev. C}\ }\textbf {\bibinfo {volume} {81}},\ \bibinfo {pages} {064607} (\bibinfo {year} {2010})}\BibitemShut {NoStop}%
\bibitem [{\citenamefont {Zhang}\ \emph {et~al.}(2021)\citenamefont {Zhang}, \citenamefont {Zhang}, \citenamefont {Li}, \citenamefont {Cheng}, \citenamefont {Liu},\ and\ \citenamefont {Zhang}}]{PhysRevC.103.024608}%
  \BibitemOpen
  \bibfield  {author} {\bibinfo {author} {\bibfnamefont {X.-R.}\ \bibnamefont {Zhang}}, \bibinfo {author} {\bibfnamefont {G.}~\bibnamefont {Zhang}}, \bibinfo {author} {\bibfnamefont {J.-J.}\ \bibnamefont {Li}}, \bibinfo {author} {\bibfnamefont {S.-H.}\ \bibnamefont {Cheng}}, \bibinfo {author} {\bibfnamefont {Z.}~\bibnamefont {Liu}}, \ and\ \bibinfo {author} {\bibfnamefont {F.-S.}\ \bibnamefont {Zhang}},\ }\href {\doibase 10.1103/PhysRevC.103.024608} {\bibfield  {journal} {\bibinfo  {journal} {Phys. Rev. C}\ }\textbf {\bibinfo {volume} {103}},\ \bibinfo {pages} {024608} (\bibinfo {year} {2021})}\BibitemShut {NoStop}%
\bibitem [{\citenamefont {Saiko}\ and\ \citenamefont {Karpov}(2019)}]{saiko2019analysis}%
  \BibitemOpen
  \bibfield  {author} {\bibinfo {author} {\bibfnamefont {V.~V.}\ \bibnamefont {Saiko}}\ and\ \bibinfo {author} {\bibfnamefont {A.~V.}\ \bibnamefont {Karpov}},\ }\href {\doibase 10.1103/PhysRevC.99.014613} {\bibfield  {journal} {\bibinfo  {journal} {Phys. Rev. C}\ }\textbf {\bibinfo {volume} {99}},\ \bibinfo {pages} {014613} (\bibinfo {year} {2019})}\BibitemShut {NoStop}%
\bibitem [{\citenamefont {Zhao}\ \emph {et~al.}(2013)\citenamefont {Zhao}, \citenamefont {Li}, \citenamefont {Wu},\ and\ \citenamefont {Zhang}}]{PhysRevC.88.044605}%
  \BibitemOpen
  \bibfield  {author} {\bibinfo {author} {\bibfnamefont {K.}~\bibnamefont {Zhao}}, \bibinfo {author} {\bibfnamefont {Z.}~\bibnamefont {Li}}, \bibinfo {author} {\bibfnamefont {X.}~\bibnamefont {Wu}}, \ and\ \bibinfo {author} {\bibfnamefont {Y.}~\bibnamefont {Zhang}},\ }\href {\doibase 10.1103/PhysRevC.88.044605} {\bibfield  {journal} {\bibinfo  {journal} {Phys. Rev. C}\ }\textbf {\bibinfo {volume} {88}},\ \bibinfo {pages} {044605} (\bibinfo {year} {2013})}\BibitemShut {NoStop}%
\bibitem [{\citenamefont {Zhang}\ \emph {et~al.}(2024{\natexlab{b}})\citenamefont {Zhang}, \citenamefont {Yao}, \citenamefont {Li}, \citenamefont {Li}, \citenamefont {Yang},\ and\ \citenamefont {Zhang}}]{zhang2024study}%
  \BibitemOpen
  \bibfield  {author} {\bibinfo {author} {\bibfnamefont {X.}~\bibnamefont {Zhang}}, \bibinfo {author} {\bibfnamefont {H.}~\bibnamefont {Yao}}, \bibinfo {author} {\bibfnamefont {C.}~\bibnamefont {Li}}, \bibinfo {author} {\bibfnamefont {T.}~\bibnamefont {Li}}, \bibinfo {author} {\bibfnamefont {Y.-X.}\ \bibnamefont {Yang}}, \ and\ \bibinfo {author} {\bibfnamefont {F.-S.}\ \bibnamefont {Zhang}},\ }\href {\doibase 10.1088/1572-9494/ad6ef4} {\bibfield  {journal} {\bibinfo  {journal} {Commun. Theor. Phys.}\ }\textbf {\bibinfo {volume} {76}},\ \bibinfo {pages} {125301} (\bibinfo {year} {2024}{\natexlab{b}})}\BibitemShut {NoStop}%
\bibitem [{\citenamefont {Feng}\ \emph {et~al.}(2024)\citenamefont {Feng}, \citenamefont {Liu}, \citenamefont {Huang}, \citenamefont {Gu}, \citenamefont {Xiao}, \citenamefont {Lei}, \citenamefont {Wang}, \citenamefont {Huang}, \citenamefont {Zhu},\ and\ \citenamefont {Su}}]{feng2024microscopic}%
  \BibitemOpen
  \bibfield  {author} {\bibinfo {author} {\bibfnamefont {Y.}~\bibnamefont {Feng}}, \bibinfo {author} {\bibfnamefont {H.}~\bibnamefont {Liu}}, \bibinfo {author} {\bibfnamefont {Y.}~\bibnamefont {Huang}}, \bibinfo {author} {\bibfnamefont {F.}~\bibnamefont {Gu}}, \bibinfo {author} {\bibfnamefont {E.}~\bibnamefont {Xiao}}, \bibinfo {author} {\bibfnamefont {X.}~\bibnamefont {Lei}}, \bibinfo {author} {\bibfnamefont {H.}~\bibnamefont {Wang}}, \bibinfo {author} {\bibfnamefont {J.}~\bibnamefont {Huang}}, \bibinfo {author} {\bibfnamefont {L.}~\bibnamefont {Zhu}}, \ and\ \bibinfo {author} {\bibfnamefont {J.}~\bibnamefont {Su}},\ }\href {\doibase 10.1103/PhysRevC.109.054604} {\bibfield  {journal} {\bibinfo  {journal} {Phys. Rev. C}\ }\textbf {\bibinfo {volume} {109}},\ \bibinfo {pages} {054604} (\bibinfo {year} {2024})}\BibitemShut {NoStop}%
\bibitem [{\citenamefont {Simenel}\ \emph {et~al.}(2020)\citenamefont {Simenel}, \citenamefont {Godbey},\ and\ \citenamefont {Umar}}]{PhysRevLett.124.212504}%
  \BibitemOpen
  \bibfield  {author} {\bibinfo {author} {\bibfnamefont {C.}~\bibnamefont {Simenel}}, \bibinfo {author} {\bibfnamefont {K.}~\bibnamefont {Godbey}}, \ and\ \bibinfo {author} {\bibfnamefont {A.~S.}\ \bibnamefont {Umar}},\ }\href {\doibase 10.1103/PhysRevLett.124.212504} {\bibfield  {journal} {\bibinfo  {journal} {Phys. Rev. Lett.}\ }\textbf {\bibinfo {volume} {124}},\ \bibinfo {pages} {212504} (\bibinfo {year} {2020})}\BibitemShut {NoStop}%
\bibitem [{\citenamefont {Yang}\ \emph {et~al.}(2025{\natexlab{b}})\citenamefont {Yang}, \citenamefont {Liao}, \citenamefont {Gao}, \citenamefont {Zhu}, \citenamefont {Su},\ and\ \citenamefont {Li}}]{yangyu}%
  \BibitemOpen
  \bibfield  {author} {\bibinfo {author} {\bibfnamefont {Y.}~\bibnamefont {Yang}}, \bibinfo {author} {\bibfnamefont {Z.}~\bibnamefont {Liao}}, \bibinfo {author} {\bibfnamefont {Z.}~\bibnamefont {Gao}}, \bibinfo {author} {\bibfnamefont {L.}~\bibnamefont {Zhu}}, \bibinfo {author} {\bibfnamefont {J.}~\bibnamefont {Su}}, \ and\ \bibinfo {author} {\bibfnamefont {C.}~\bibnamefont {Li}},\ }\href {\doibase 10.1103/PhysRevC.111.024602} {\bibfield  {journal} {\bibinfo  {journal} {Phys. Rev. C}\ }\textbf {\bibinfo {volume} {111}},\ \bibinfo {pages} {024602} (\bibinfo {year} {2025}{\natexlab{b}})}\BibitemShut {NoStop}%
\bibitem [{\citenamefont {Simenel}(2010)}]{simenel2010particle}%
  \BibitemOpen
  \bibfield  {author} {\bibinfo {author} {\bibfnamefont {C.}~\bibnamefont {Simenel}},\ }\href {\doibase 10.1103/PhysRevLett.105.192701} {\bibfield  {journal} {\bibinfo  {journal} {Phys. Rev. Lett.}\ }\textbf {\bibinfo {volume} {105}},\ \bibinfo {pages} {192701} (\bibinfo {year} {2010})}\BibitemShut {NoStop}%
\bibitem [{\citenamefont {Sekizawa}\ and\ \citenamefont {Yabana}(2013)}]{sekizawa2013time}%
  \BibitemOpen
  \bibfield  {author} {\bibinfo {author} {\bibfnamefont {K.}~\bibnamefont {Sekizawa}}\ and\ \bibinfo {author} {\bibfnamefont {K.}~\bibnamefont {Yabana}},\ }\href {\doibase 10.1103/PhysRevC.88.014614} {\bibfield  {journal} {\bibinfo  {journal} {Phys. Rev. C}\ }\textbf {\bibinfo {volume} {88}},\ \bibinfo {pages} {014614} (\bibinfo {year} {2013})}\BibitemShut {NoStop}%
\bibitem [{\citenamefont {Jiang}\ and\ \citenamefont {Wang}(2020)}]{jiang2020probing}%
  \BibitemOpen
  \bibfield  {author} {\bibinfo {author} {\bibfnamefont {X.}~\bibnamefont {Jiang}}\ and\ \bibinfo {author} {\bibfnamefont {N.}~\bibnamefont {Wang}},\ }\href {\doibase 10.1103/PhysRevC.101.014604} {\bibfield  {journal} {\bibinfo  {journal} {Phys. Rev. C}\ }\textbf {\bibinfo {volume} {101}},\ \bibinfo {pages} {014604} (\bibinfo {year} {2020})}\BibitemShut {NoStop}%
\bibitem [{\citenamefont {Zhang}\ \emph {et~al.}(2024{\natexlab{c}})\citenamefont {Zhang}, \citenamefont {Vretenar}, \citenamefont {Nik\ifmmode \check{s}\else \v{s}\fi{}i\ifmmode~\acute{c}\else \'{c}\fi{}}, \citenamefont {Zhao},\ and\ \citenamefont {Meng}}]{PhysRevC.109.024614}%
  \BibitemOpen
  \bibfield  {author} {\bibinfo {author} {\bibfnamefont {D.~D.}\ \bibnamefont {Zhang}}, \bibinfo {author} {\bibfnamefont {D.}~\bibnamefont {Vretenar}}, \bibinfo {author} {\bibfnamefont {T.}~\bibnamefont {Nik\ifmmode \check{s}\else \v{s}\fi{}i\ifmmode~\acute{c}\else \'{c}\fi{}}}, \bibinfo {author} {\bibfnamefont {P.~W.}\ \bibnamefont {Zhao}}, \ and\ \bibinfo {author} {\bibfnamefont {J.}~\bibnamefont {Meng}},\ }\href {\doibase 10.1103/PhysRevC.109.024614} {\bibfield  {journal} {\bibinfo  {journal} {Phys. Rev. C}\ }\textbf {\bibinfo {volume} {109}},\ \bibinfo {pages} {024614} (\bibinfo {year} {2024}{\natexlab{c}})}\BibitemShut {NoStop}%
\bibitem [{\citenamefont {Golabek}\ and\ \citenamefont {Simenel}(2009)}]{PhysRevLett.103.042701}%
  \BibitemOpen
  \bibfield  {author} {\bibinfo {author} {\bibfnamefont {C.}~\bibnamefont {Golabek}}\ and\ \bibinfo {author} {\bibfnamefont {C.}~\bibnamefont {Simenel}},\ }\href {\doibase 10.1103/PhysRevLett.103.042701} {\bibfield  {journal} {\bibinfo  {journal} {Phys. Rev. Lett.}\ }\textbf {\bibinfo {volume} {103}},\ \bibinfo {pages} {042701} (\bibinfo {year} {2009})}\BibitemShut {NoStop}%
\bibitem [{\citenamefont {Zhang}\ \emph {et~al.}(2024{\natexlab{d}})\citenamefont {Zhang}, \citenamefont {Li}, \citenamefont {Vretenar}, \citenamefont {Nik\ifmmode \check{s}\else \v{s}\fi{}i\ifmmode~\acute{c}\else \'{c}\fi{}}, \citenamefont {Ren}, \citenamefont {Zhao},\ and\ \citenamefont {Meng}}]{PhysRevC.109.024316}%
  \BibitemOpen
  \bibfield  {author} {\bibinfo {author} {\bibfnamefont {D.~D.}\ \bibnamefont {Zhang}}, \bibinfo {author} {\bibfnamefont {B.}~\bibnamefont {Li}}, \bibinfo {author} {\bibfnamefont {D.}~\bibnamefont {Vretenar}}, \bibinfo {author} {\bibfnamefont {T.}~\bibnamefont {Nik\ifmmode \check{s}\else \v{s}\fi{}i\ifmmode~\acute{c}\else \'{c}\fi{}}}, \bibinfo {author} {\bibfnamefont {Z.~X.}\ \bibnamefont {Ren}}, \bibinfo {author} {\bibfnamefont {P.~W.}\ \bibnamefont {Zhao}}, \ and\ \bibinfo {author} {\bibfnamefont {J.}~\bibnamefont {Meng}},\ }\href {\doibase 10.1103/PhysRevC.109.024316} {\bibfield  {journal} {\bibinfo  {journal} {Phys. Rev. C}\ }\textbf {\bibinfo {volume} {109}},\ \bibinfo {pages} {024316} (\bibinfo {year} {2024}{\natexlab{d}})}\BibitemShut {NoStop}%
\bibitem [{\citenamefont {Charity}(2010)}]{PhysRevC.82.014610}%
  \BibitemOpen
  \bibfield  {author} {\bibinfo {author} {\bibfnamefont {R.~J.}\ \bibnamefont {Charity}},\ }\href {\doibase 10.1103/PhysRevC.82.014610} {\bibfield  {journal} {\bibinfo  {journal} {Phys. Rev. C}\ }\textbf {\bibinfo {volume} {82}},\ \bibinfo {pages} {014610} (\bibinfo {year} {2010})}\BibitemShut {NoStop}%
\bibitem [{\citenamefont {Wang}\ \emph {et~al.}(2021)\citenamefont {Wang}, \citenamefont {Huang}, \citenamefont {Kondev}, \citenamefont {Audi},\ and\ \citenamefont {Naimi}}]{Wang_2021}%
  \BibitemOpen
  \bibfield  {author} {\bibinfo {author} {\bibfnamefont {M.}~\bibnamefont {Wang}}, \bibinfo {author} {\bibfnamefont {W.}~\bibnamefont {Huang}}, \bibinfo {author} {\bibfnamefont {F.}~\bibnamefont {Kondev}}, \bibinfo {author} {\bibfnamefont {G.}~\bibnamefont {Audi}}, \ and\ \bibinfo {author} {\bibfnamefont {S.}~\bibnamefont {Naimi}},\ }\href {\doibase 10.1088/1674-1137/abddaf} {\bibfield  {journal} {\bibinfo  {journal} {Chin. Phys. C}\ }\textbf {\bibinfo {volume} {45}},\ \bibinfo {pages} {030003} (\bibinfo {year} {2021})}\BibitemShut {NoStop}%
\bibitem [{\citenamefont {Gao}\ \emph {et~al.}(2021)\citenamefont {Gao}, \citenamefont {Wang}, \citenamefont {L{\"u}}, \citenamefont {Li}, \citenamefont {Shen},\ and\ \citenamefont {Liu}}]{gao2021machine}%
  \BibitemOpen
  \bibfield  {author} {\bibinfo {author} {\bibfnamefont {Z.-P.}\ \bibnamefont {Gao}}, \bibinfo {author} {\bibfnamefont {Y.-J.}\ \bibnamefont {Wang}}, \bibinfo {author} {\bibfnamefont {H.-L.}\ \bibnamefont {L{\"u}}}, \bibinfo {author} {\bibfnamefont {Q.-F.}\ \bibnamefont {Li}}, \bibinfo {author} {\bibfnamefont {C.-W.}\ \bibnamefont {Shen}}, \ and\ \bibinfo {author} {\bibfnamefont {L.}~\bibnamefont {Liu}},\ }\href {\doibase https://doi.org/10.1007/s41365-021-00956-1} {\bibfield  {journal} {\bibinfo  {journal} {Nucl. Sci. Tech.}\ }\textbf {\bibinfo {volume} {32}},\ \bibinfo {pages} {109} (\bibinfo {year} {2021})}\BibitemShut {NoStop}%
\bibitem [{\citenamefont {Devaraja}\ \emph {et~al.}(2020)\citenamefont {Devaraja}, \citenamefont {Heinz}, \citenamefont {Ackermann}, \citenamefont {G{\"o}bel}, \citenamefont {He{\ss}berger}, \citenamefont {Hofmann}, \citenamefont {Maurer}, \citenamefont {M{\"u}nzenberg}, \citenamefont {Popeko},\ and\ \citenamefont {Yeremin}}]{devaraja2020new}%
  \BibitemOpen
  \bibfield  {author} {\bibinfo {author} {\bibfnamefont {H.}~\bibnamefont {Devaraja}}, \bibinfo {author} {\bibfnamefont {S.}~\bibnamefont {Heinz}}, \bibinfo {author} {\bibfnamefont {D.}~\bibnamefont {Ackermann}}, \bibinfo {author} {\bibfnamefont {T.}~\bibnamefont {G{\"o}bel}}, \bibinfo {author} {\bibfnamefont {F.}~\bibnamefont {He{\ss}berger}}, \bibinfo {author} {\bibfnamefont {S.}~\bibnamefont {Hofmann}}, \bibinfo {author} {\bibfnamefont {J.}~\bibnamefont {Maurer}}, \bibinfo {author} {\bibfnamefont {G.}~\bibnamefont {M{\"u}nzenberg}}, \bibinfo {author} {\bibfnamefont {A.}~\bibnamefont {Popeko}}, \ and\ \bibinfo {author} {\bibfnamefont {A.}~\bibnamefont {Yeremin}},\ }\href {\doibase https://doi.org/10.1140/epja/s10050-020-00229-2} {\bibfield  {journal} {\bibinfo  {journal} {Eur. Phys. J. A}\ }\textbf {\bibinfo {volume} {56}},\ \bibinfo {pages} {224} (\bibinfo {year} {2020})}\BibitemShut {NoStop}%
\bibitem [{\citenamefont {Devaraja}\ \emph {et~al.}(2015)\citenamefont {Devaraja}, \citenamefont {Heinz}, \citenamefont {Beliuskina}, \citenamefont {Comas}, \citenamefont {Hofmann}, \citenamefont {Hornung}, \citenamefont {M{\"u}nzenberg}, \citenamefont {Nishio}, \citenamefont {Ackermann}, \citenamefont {Gambhir} \emph {et~al.}}]{devaraja2015observation}%
  \BibitemOpen
  \bibfield  {author} {\bibinfo {author} {\bibfnamefont {H.}~\bibnamefont {Devaraja}}, \bibinfo {author} {\bibfnamefont {S.}~\bibnamefont {Heinz}}, \bibinfo {author} {\bibfnamefont {O.}~\bibnamefont {Beliuskina}}, \bibinfo {author} {\bibfnamefont {V.}~\bibnamefont {Comas}}, \bibinfo {author} {\bibfnamefont {S.}~\bibnamefont {Hofmann}}, \bibinfo {author} {\bibfnamefont {C.}~\bibnamefont {Hornung}}, \bibinfo {author} {\bibfnamefont {G.}~\bibnamefont {M{\"u}nzenberg}}, \bibinfo {author} {\bibfnamefont {K.}~\bibnamefont {Nishio}}, \bibinfo {author} {\bibfnamefont {D.}~\bibnamefont {Ackermann}}, \bibinfo {author} {\bibfnamefont {Y.}~\bibnamefont {Gambhir}},  \emph {et~al.},\ }\href {\doibase https://doi.org/10.1016/j.physletb.2015.07.006} {\bibfield  {journal} {\bibinfo  {journal} {Phys. Lett. B}\ }\textbf {\bibinfo {volume} {748}},\ \bibinfo {pages} {199} (\bibinfo {year} {2015})}\BibitemShut {NoStop}%
\bibitem [{\citenamefont {Devaraja}\ \emph {et~al.}(2019)\citenamefont {Devaraja}, \citenamefont {Heinz}, \citenamefont {Beliuskina}, \citenamefont {Hofmann}, \citenamefont {Hornung}, \citenamefont {M{\"u}nzenberg}, \citenamefont {Ackermann}, \citenamefont {Gupta}, \citenamefont {Gambhir}, \citenamefont {Henderson} \emph {et~al.}}]{devaraja2019population}%
  \BibitemOpen
  \bibfield  {author} {\bibinfo {author} {\bibfnamefont {H.}~\bibnamefont {Devaraja}}, \bibinfo {author} {\bibfnamefont {S.}~\bibnamefont {Heinz}}, \bibinfo {author} {\bibfnamefont {O.}~\bibnamefont {Beliuskina}}, \bibinfo {author} {\bibfnamefont {S.}~\bibnamefont {Hofmann}}, \bibinfo {author} {\bibfnamefont {C.}~\bibnamefont {Hornung}}, \bibinfo {author} {\bibfnamefont {G.}~\bibnamefont {M{\"u}nzenberg}}, \bibinfo {author} {\bibfnamefont {D.}~\bibnamefont {Ackermann}}, \bibinfo {author} {\bibfnamefont {M.}~\bibnamefont {Gupta}}, \bibinfo {author} {\bibfnamefont {Y.}~\bibnamefont {Gambhir}}, \bibinfo {author} {\bibfnamefont {R.}~\bibnamefont {Henderson}},  \emph {et~al.},\ }\href {\doibase https://doi.org/10.1140/epja/i2019-12696-3} {\bibfield  {journal} {\bibinfo  {journal} {The European Physical Journal A}\ }\textbf {\bibinfo {volume} {55}},\ \bibinfo {pages} {25} (\bibinfo {year} {2019})}\BibitemShut {NoStop}%
\bibitem [{\citenamefont {Di~Nitto}\ \emph {et~al.}(2018)\citenamefont {Di~Nitto}, \citenamefont {Khuyagbaatar}, \citenamefont {Ackermann}, \citenamefont {Andersson}, \citenamefont {Badura}, \citenamefont {Block}, \citenamefont {Brand}, \citenamefont {Conrad}, \citenamefont {Cox}, \citenamefont {D{\"u}llmann} \emph {et~al.}}]{di2018study}%
  \BibitemOpen
  \bibfield  {author} {\bibinfo {author} {\bibfnamefont {A.}~\bibnamefont {Di~Nitto}}, \bibinfo {author} {\bibfnamefont {J.}~\bibnamefont {Khuyagbaatar}}, \bibinfo {author} {\bibfnamefont {D.}~\bibnamefont {Ackermann}}, \bibinfo {author} {\bibfnamefont {L.-L.}\ \bibnamefont {Andersson}}, \bibinfo {author} {\bibfnamefont {E.}~\bibnamefont {Badura}}, \bibinfo {author} {\bibfnamefont {M.}~\bibnamefont {Block}}, \bibinfo {author} {\bibfnamefont {H.}~\bibnamefont {Brand}}, \bibinfo {author} {\bibfnamefont {I.}~\bibnamefont {Conrad}}, \bibinfo {author} {\bibfnamefont {D.}~\bibnamefont {Cox}}, \bibinfo {author} {\bibfnamefont {C.~E.}\ \bibnamefont {D{\"u}llmann}},  \emph {et~al.},\ }\href {\doibase https://doi.org/10.1016/j.physletb.2018.07.058} {\bibfield  {journal} {\bibinfo  {journal} {Phys. Lett. B}\ }\textbf {\bibinfo {volume} {784}},\ \bibinfo {pages} {199} (\bibinfo {year} {2018})}\BibitemShut {NoStop}%
\bibitem [{bou()}]{boundary}%
  \BibitemOpen
  \href {https://www-nds.iaea.org/relnsd/vcharthtml/VChartHTML.html} {\bibinfo  {journal} {www-nds.iaea.org/relnsd/vcharthtml/VChartHTML.html}\ }\BibitemShut {NoStop}%
\end{thebibliography}
%

\end{document}